\documentclass[fleqn,usenatbib]{mnras}

\usepackage{newtxtext,newtxmath}

\usepackage[T1]{fontenc}
\usepackage{ae,aecompl}

\usepackage{graphicx}	
\usepackage{amsmath}	
\usepackage{amssymb}	

\newcommand{\ec}{$\eta$ Carinae}
\newcommand{\kmps}{\ensuremath{\mathrm{km~s^{-1}}}}
\newcommand{\Msun}{\ensuremath{\mathrm{M_\odot}}}

\newcommand{\Msunpyr}{\ensuremath{\mathrm{M_\odot~yr^{-1}}}}

\title[iPTF14hls as a variable hyper-wind]{iPTF14hls as a variable hyper-wind from a very massive star}

\author[T. J. Moriya et al.]{
Takashi J. Moriya,$^{1,2}$\thanks{E-mail: takashi.moriya@nao.ac.jp (TJM)}
Paolo A. Mazzali,$^{3,4}$
and Elena Pian$^5$
\\
$^{1}$ National Astronomical Observatory of Japan, National Institutes of Natural Sciences, 2-21-1 Osawa, Mitaka, Tokyo 181-8588, Japan\\
$^{2}$ School of Physics and Astronomy, Faculty of Science, Monash University, Clayton, VIC 3800, Australia\\
$^{3}$ Astrophysics Research Institute, Liverpool John Moores University, IC2, Liverpool Science Park, 146 Brownlow Hill, Liverpool L3 5RF, UK\\
$^{4}$ Max-Planck Institute for Astrophysics, Karl-Schwarzschild-Stra{\ss}e 1, 85748 Garching, Germany \\
$^{5}$ INAF, Astrophysics and Space Science Observatory, via P. Gobetti 101, 40129 Bologna, Italy 
}

\date{Accepted 2019 October 18. Received 2019 October 16; in original form 2019 September 04}

\pubyear{2019}

\begin{document}
\label{firstpage}
\pagerange{\pageref{firstpage}--\pageref{lastpage}}
\maketitle

\begin{abstract}
The origin of iPTF14hls, which had Type~IIP supernova-like spectra but kept bright for almost two years with little spectral evolution, is still unclear. We here propose that iPTF14hls was not a sudden outburst like supernovae but rather a long-term outflow similar to stellar winds. The properties of iPTF14hls, which are at odds with a supernova scenario, become natural when interpreted as a stellar wind with variable mass-loss rate. Based on the wind hypothesis, we estimate the mass-loss rates of iPTF14hls in the bright phase. We find that the instantaneous mass-loss rate of iPTF14hls during the 2-year bright phase was more than a few \Msunpyr\ (``hyper-wind'') and it reached as much as 10~\Msunpyr\ . The total mass lost over two years was about 10~\Msun. Interestingly, we find that the light curve of iPTF14hls has a very
similar shape to that of \ec\ during the Great Eruption, which also experienced a similar but less extreme brightening accompanied by extraordinary mass loss, shedding more than 10~\Msun\ in 10~years. The progenitor of iPTF14hls is less than 150~\Msun\ if it still exists, which is similar to \ec. The two phenomena may be related to a continuum-driven extreme wind from very massive stars.
\end{abstract}




\section{Introduction}
Our Universe is dynamic in time. We regularly observe brightening and fading of stars. Supernovae (SNe) are among the brightest variable phenomena in our Universe.
iPTF14hls became as bright as SNe and showed spectra typical of SNe originating from the collapse of massive stars \citep{arcavi2017,sollerman2019latephase}.
iPTF14hls was, however, very bright for almost 2 years, while SNe showing similar spectra typically last for only 100~days and reach a much lower luminosity \citep[e.g.,][]{anderson2014}.
iPTF14hls rose in luminosity over about three months and remained very luminous for
more than one year. Its spectra were dominated by hydrogen lines as in Type~IIP
SNe \citep{filippenko1997} and remained similar for almost 2 years \citep{arcavi2017}. During this
period, changes in line velocity and photospheric radii were subtle, unlike those
observed in SNe \citep{arcavi2017}.

In SNe, mass ejection occurs  over a small timescale, and we observe the ejecta as they expand. Therefore, we see deeper and slower parts of the ejecta as time goes
by \citep{mazzali04eo,kasen2009iip,bersten2011iip}. This is tracked by the evolution of the photospheric velocity, which decreases with time. This velocity evolution was not seen in iPTF14hls.  Thus, iPTF14hls needs to have unusual physical conditions for a SN and it was claimed to be a peculiar SN in previous
studies \citep[e.g.,][]{dessart2018mag,woosley2018,chugai2018,soker2018comenvjet,wang2018fallback,liu2019fallback,gofman2019binary}.

A lack of evolution of the photospheric properties is not expected in a mass
ejection that takes place in a short timescale, but is natural for a
steady-state situation, such as mass loss over a long timescale. An
obvious example of this kind is a stellar wind. We do observe constant line
velocities and photospheric properties in stellar winds, where mass ejection
from the stellar surface is continuous. What makes iPTF14hls peculiar as a
short-time explosive phenomenon becomes natural if it was a long-term continuous
outflow. In this paper, we propose that iPTF14hls is related to an extreme,
continuous, variable, long-term time-resolved mass outflow from a massive star rather than
a violent, short-term mass ejection like a SN.
We call such an extreme mass outflow from massive stars, which exceeds $1~\Msunpyr$ in the case of iPTF14hls as we shall see, a ``hyper-wind.''

\begin{figure}
\includegraphics[width=\columnwidth]{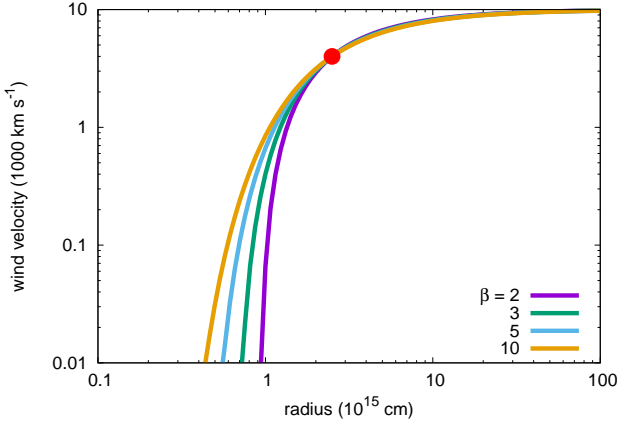} 
\caption{
Possible wind velocity structure in iPTF14hls.
}
\label{fig:wind_structure}
\end{figure}

\section{iPTF14hls as a stellar wind}\label{sec:wind}
We first discuss the velocity structure of the wind in iPTF14hls leading to its extremely slow spectral evolution. \citet{arcavi2017} estimated that the velocity measured from the P-Cygni profile of Fe~\textsc{ii} $\lambda 5169$ is $\simeq 4000~\kmps$, which can be considered as the velocity at the photosphere. The velocities measured from the hydrogen lines are faster, i.e., $8000-6000~\kmps$ for H$\alpha$ and $7000-5000~\kmps$ for H$\beta$. The absorption in the hydrogen lines extends to $\simeq 10000~\kmps$ and, therefore, the maximum velocity reached by the wind is likely around 10000~\kmps. We interpret that 10000~\kmps\ is the terminal velocity of the wind forming iPTF14hls and the lower velocities are found as the line formation occurs in the wind accelerated towards the terminal velocity.

The acceleration of the wind is often approximated by using the so-called $\beta$ velocity law, i.e.,
\begin{equation}
    V_\mathrm{wind}(r) = V_\infty \left(1-\frac{R_0}{r} \right)^\beta, \label{eq:betalaw}
\end{equation}
where $V_\infty = 10000~\kmps$ is the terminal wind velocity and $R_0$ is the radius where the wind is launched. $\beta$ represents how fast the wind is accelerated. As $\beta$ increases, the wind is more slowly accelerated. Extended stars such as red supergiants (RSGs) are known to have a slow wind acceleration with $\beta\gtrsim 2$ \citep[e.g.,][]{bennett2010rsgwind}. Because iPTF14hls has a large photospheric radius ($R_\mathrm{ph}\simeq 2.5\times 10^{15}~\mathrm{cm}$) as shown in the following Section~\ref{sec:masslossrates}, the wind acceleration in iPTF14hls is likely slow and $\beta$ could be at least as large as those found in RSGs.

Taking $R_\mathrm{ph}\simeq 2.5\times 10^{15}~\mathrm{cm}$ and $v_\mathrm{ph}\simeq 4000~\kmps$, we can calibrate Eq.~(\ref{eq:betalaw}) for given $\beta$ and see the wind velocity structure in iPTF14hls. Although the wind does not necessarily follow the $\beta$ law, we can have a general idea of the wind structure in this way.

The wind velocity and density structure estimated with the above method is shown in Fig.~\ref{fig:wind_structure}. Although the wind velocity at the photosphere ($\simeq 2.5\times 10^{15}~\mathrm{cm}$) is 4000~\kmps, the terminal wind velocity is achieved well above the photosphere. If this velocity structure does not change much, we can observe P-Cygni profiles that do not change much over time because this velocity structure is fixed in the radius. This explains the spectra without apparent velocity change in iPTF14hls which is not expected in expanding ejecta.

The wind launching radius $R_0$ strongly depends on the assumed $\beta$ if we adopt the $\beta$ velocity law (Fig.~\ref{fig:wind_structure}). They are
$(5-10)\times 10^{14}~\mathrm{cm}$ with $\beta\simeq 2-5$.
The progenitor radius of iPTF14hls is, therefore, likely around $(5-10)\times 10^{14}~\mathrm{cm}$.

\begin{figure}
\includegraphics[width=\columnwidth]{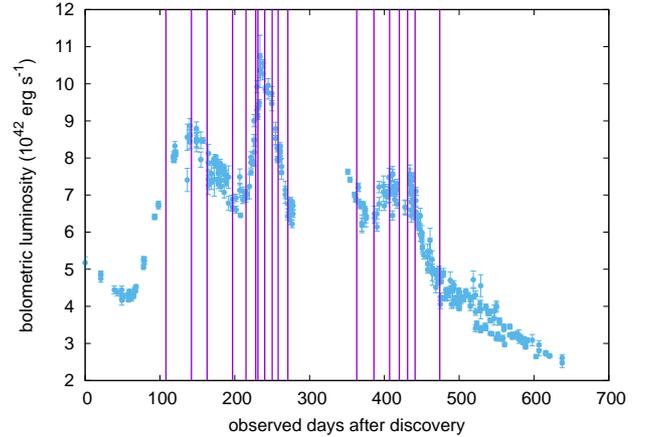} 
\caption{
Bolometric LC of iPTF14hls from \citet{sollerman2019latephase}. The epochs of the 18 spectra we selected for our analysis are marked by vertical purple lines.  All spectra (reported in Fig. \ref{fig:bbfits}) were fitted with a blackbody and the fitting parameters were used to estimate the photospheric properties.
}
\label{fig:fittedepoch}
\end{figure}

\section{Mass-loss rates of the progenitor}\label{sec:masslossrates}
If iPTF14hls is a stellar wind, the next important question is the mass-loss rates of the progenitor causing iPTF14hls. To estimate the mass-loss rates of iPTF14hls, we first 
evaluate photospheric radii and temperatures of iPTF14hls by fitting blackbody functions to selected spectra. The epochs of the spectra when we performed the blackbody fitting are shown in Fig.~\ref{fig:fittedepoch}. We took the spectra at these epochs from WISeREP\footnote{\url{https://wiserep.weizmann.ac.il}} \citep{yaron2012wiserep}. These spectra are reported by \citet{arcavi2017}. The spectra in the database were flux calibrated based on the photometry \citep{arcavi2017}. All the flux calibrated spectra we used are shown in Appendix~\ref{app:bbfit} with photometry.

Our blackbody fitting was performed by taking line blanketing in the blue and ultraviolet into account. Even if photons are emitted with the blackbody energy distribution from the photosphere, photons can be absorbed and re-emitted above the photosphere and the spectra do not keep the original blackbody shape. This line blanketing mostly absorbs blue and ultraviolet photons and they are re-emitted in redder wavelengths. Thus, this effect changes the original blackbody spectra especially in the blue and ultraviolet. Thus, we fit the spectra by matching the redder part of the continuum spectra to the blackbody function and allowing flux excess in the blue and ultraviolet. The previous studies obtaining the blackbody properties of iPTF14hls \citep{arcavi2017,sollerman2019latephase} simply fitted photometry with the blackbody function and did not take the line blanketing effect into account.

The results of the blackbody fitting with errors are presented in Appendix~\ref{app:bbfit}. Fig.~\ref{fig:radius_temperature} shows the estimated photospheric radii and temperatures. There is some difference between our estimates and the results mentioned in previous works by \citet{arcavi2017,sollerman2019latephase} caused by the line blanketing. In particular, we found almost constant photospheric radii ($(2-3)\times 10^{15}~\mathrm{cm}$) and slightly higher photospheric temperatures ($6250-7500~\mathrm{K}$) than the previous estimates ($5000-6000~\mathrm{K}$). The temperatures we derive are consistent with the recombination temperature of ionised hydrogen and support the idea that the spectra are formed at the electron scattering photosphere of the hydrogen-rich wind, as also suggested by the increasingly strong H$\alpha$ emission component. This causes the spectra of iPTF14hls to resemble those of Type\,IIP SNe.

Based on our measurements of the photospheric radii and temperatures, we estimate the mass-loss rate of iPTF14hls. The mass-loss rate at the photosphere is expressed as
\begin{equation}
\dot{M} = 4\pi R_\mathrm{ph}^2 \rho_\mathrm{ph} V_\mathrm{ph},    
\end{equation}
where
$R_\mathrm{ph}$ is the photospheric radius, $\rho_\mathrm{ph}$ is the density at photosphere, and $v_\mathrm{ph}$ is the photospheric velocity. In order to form the P-Cygni profiles of hydrogen lines as observed in iPTF14hls, the electron number density is required to be $\simeq (8\pm1)\times 10^{9}~\mathrm{cm^{-3}}$ when the photospheric temperature is $6000-7000~\mathrm{K}$ as found in iPTF14hls \citep{eastman1996,owocki2016eccdw}. Assuming a solar composition, the corresponding photospheric density is $\rho_\mathrm{ph}\simeq (1.8\pm 0.3)\times 10^{-14}~\mathrm{g~cm^{-3}}$. The velocity at the photosphere is $V_\mathrm{ph}\simeq 4000~\kmps$ throughout the peak phase based on Fe~\textsc{ii} $\lambda 5169$ line measurements \citep{arcavi2017,sollerman2019latephase}. We use the photospheric radii presented in Fig.~\ref{fig:radius_temperature}.

\begin{figure}
\includegraphics[width=\columnwidth]{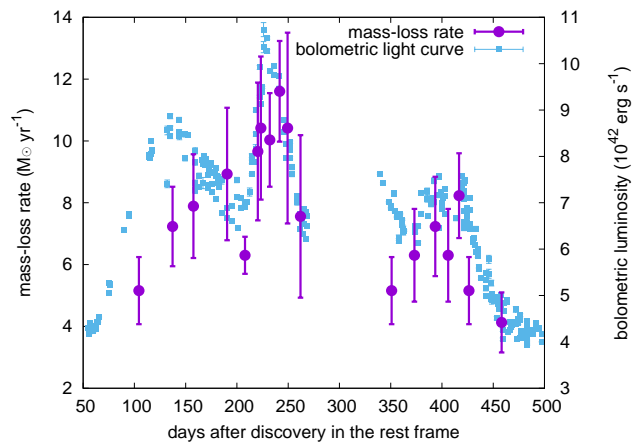}
\caption{
Estimated mass-loss rate history of iPTF14hls (left axis). The bolometric LC of iPTF14hls \citep{sollerman2019latephase} is overplotted for comparison (right axis).
}
\label{fig:masslossrate}
\end{figure}

The estimated mass-loss rate history of iPTF14hls is presented in Fig.~\ref{fig:masslossrate}. The estimated mass-loss rates exceed 1~\Msunpyr\ and they are extremely large (``hyper-wind''). 
The mass-loss rate is estimated to be $\simeq 9~\Msunpyr$ on average from 100 days to 260 days and about 4~\Msun\ are lost in this period. Initially, up to $\sim 150$~days, the rise in
luminosity precedes that of the mass-loss rate. Simultaneously, the temperature decreases. This indicates an initial expansion, as traced by the increase in radius. This phase likely represents the onset of the high mass-loss episode. Although the mass-loss rate starts to increase when the luminosity also does, the first peak in the mass-loss rate at the photospheric radius is reached some 2 months after the first luminosity peak. This is the time it takes gas moving at 4000~\kmps\ to reach the photospheric radius ($2.5 \times 10^{15}~\mathrm{cm}$). Thereafter, the mass-loss rate follows the behaviour of the light curve (LC) -- when the LC gets brighter, the mass-loss rates become larger, and vice-versa. Several peaks in the mass-loss rates can be seen, some sharper and some smoother. In particular, a sharp peak is seen at $220-250$ days, when instantaneous mass-loss rates in excess of 10~\Msunpyr\ are reached. The radius responds to these changes in mass-loss rates, but the temperature remains roughly constant, indicating an optically thick regime. In the second observing season after $\sim\,350~\mathrm{days}$, the mass-loss rate becomes smaller as iPTF14hls gets fainter. The terminal velocity of the outflow probably decreases somewhat, as shown by the slow evolution of the observed velocity of both the  H$\alpha$ and H$\beta$ P-Cygni absorption components. The photospheric radius also starts to decrease, suggesting that the hydrogen-dominated gas is progressively recombining. This is similar to the situation of a Type\,IIP SN at the end of the plateau phase, and is confirmed by the increasing importance of H$\alpha$ emission with respect to overall luminosity.

Between 350 and 460 days the average mass-loss rate is about 6~\Msunpyr, and the total mass lost during this period is about 2~\Msun. The total mass lost in 2 years may be close to 10~\Msun.
Because the mass that is ejected is accelerated to a terminal velocity of 10000~\kmps, the total kinetic energy of the outflow may be around $10^{52}~\mathrm{erg}$ -- much higher than the standard SN explosion energy of $10^{51}~\mathrm{erg}$ \citep{pejcha2015}.

\begin{figure}
\includegraphics[width=\columnwidth]{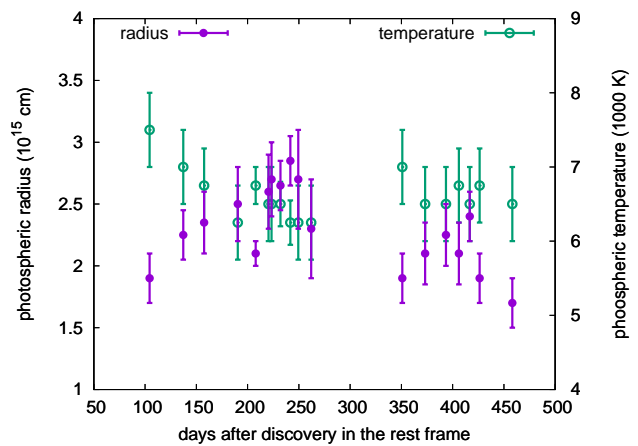} 
\caption{
Photospheric radii and temperatures of iPTF14hls. They are derived by fitting a blackbody function to the observed spectra with line blanketing in mind.
}
\label{fig:radius_temperature}
\end{figure}

\section{iPTF14hls and the ``Great Eruption''}\label{sec:greateruption}
The huge mass-loss rates exceeding 1~\Msunpyr, accompanied by significant stellar brightening in iPTF14hls are reminiscent of
\ec\ \citep{davidson1997ecar}. \ec\ suddenly became bright in 1838 and it maintained a high luminosity for about
10~years \citep{fernandez-lajus2009etacarlightcurve}.
During this period known as the ``Great Eruption'', \ec\ ejected a large amount of mass, which we currently observe as the nebula called the ``Homunculus'' \citep{davidson1997ecar}.
The mass-loss rate during the Great Eruption is estimated to have been more than 1~\Msunpyr \citep{morris1999,smith2013ec,smith2006homu}.
Interestingly, the spectra of \ec\ during the Great Eruption obtained by observing its light echoes have similar hydrogen P-Cygni profiles and an indication of a high velocity component in
hydrogen ($\simeq 10000~\kmps$) as in iPTF14hls \citep{rest2012echo,prieto2014ececho,smith2018highvec}.

\begin{figure}
\includegraphics[width=\columnwidth]{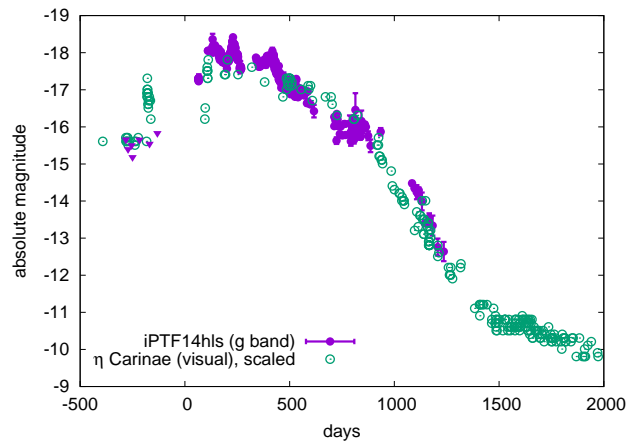} 
\caption{
Comparison of the LCs of iPTF14hls \citep{arcavi2017} and $\eta$ Carinae at the Great Eruption \citep{fernandez-lajus2009etacarlightcurve}. The LC of $\eta$ Carinae is scaled by a factor of 0.15 in time and 100 (5~magnitudes) in brightness. The triangles are the unfiltered upper limits from Catalina Sky Survey \citep{arcavi2017}.
}
\label{fig:lightcurve}
\end{figure}

If iPTF14hls is a similar mass-loss event to the Great Eruption of \ec, the progenitor may not have disappeared. Its last observed luminosity, some 3 years
after discovery, was $\approx 5\times 10^6~\mathrm{L_\odot}$ \citep{sollerman2019latephase}. The progenitor mass of iPTF14hls, therefore, must be smaller than 150~\Msun\ if it still
exists \citep[e.g.,][]{ekstrom2012,georgy2013,szecsi2015lowmetal,yoon2012popiii}. This mass is similar to (or slightly larger than) the mass of \ec\ ($\sim 100~\Msun$) \citep{davidson1997ecar}.
The mass-loss and brightening mechanisms of \ec\ are not understood well, but we found that there is a striking similarity between the LC of iPTF14hls and that of \ec\ during the
Great Eruption, when the latter is scaled in time and luminosity (Fig.~\ref{fig:lightcurve}). iPTF14hls was 100 times brighter and 6.5 times shorter-lived than the Great Eruption. The similarity may indicate that iPTF14hls and \ec\ share a common wind driving mechanism but with a different timescale, presumably because of the difference in the luminosity. The Great Eruption has indeed been suggested to be a continuum-driven wind rather than a short-term mass ejection \citep[e.g.,][]{owocki2016eccdw,davidson1987}.
Both iPTF14hls and \ec\ during the Great Eruption were well above their Eddington luminosity in the bright phases and an optically-thick continuum-driven wind can be
triggered in both cases, resulting in the very high mass-loss rates and the ionization of the material ejected \citep[e.g.,][]{vanmarle2008supewind}.

If the iPTF14hls progenitor survived and became as luminous as \ec, iPTF14hls should have been settled at around $27.5$~magnitudes in optical by now given the rapid photometric decline rate observed in early 2018 \citep{sollerman2019latephase}. Future observations by \textit{Hubble Space Telescope} or \textit{James Webb Space Telescope} will be able to judge whether there is a surviving massive star at the location of iPTF14hls.

\section{Discussion}

\subsection{Luminosity source of iPTF14hls}
\subsubsection{Early phase luminosity}
We support the interpretation that iPTF14hls was a continuous, long-lasting outflow caused by a sudden re-brightening of a very massive star. However, its spectrum looked like that of a Type~IIP SN. This is not a problem, as long as the conditions in the outflow (density, temperature) are similar to those of the line-forming region of a Type~IIP SN, where hydrogen recombination causes a sudden jump in opacity and electron scattering is the main source of opacity below that region. One of the diagnostics of hydrogen recombination is the dominance of the emission component over the absorption one in the H$\alpha$ P-Cygni profile. This is seen in Type~IIP SNe as time advances, both in the plateau and in the later radioactive tail phase, but the same is seen in iPTF14hls. We therefore use the example of Type~IIP SNe to estimate whether recombination radiation can be the main method of reprocessing the stellar radiation that makes iPTF14hls bright, just as in Type~IIP SNe.

\begin{figure*}
\center
\includegraphics[width=1.4\columnwidth]{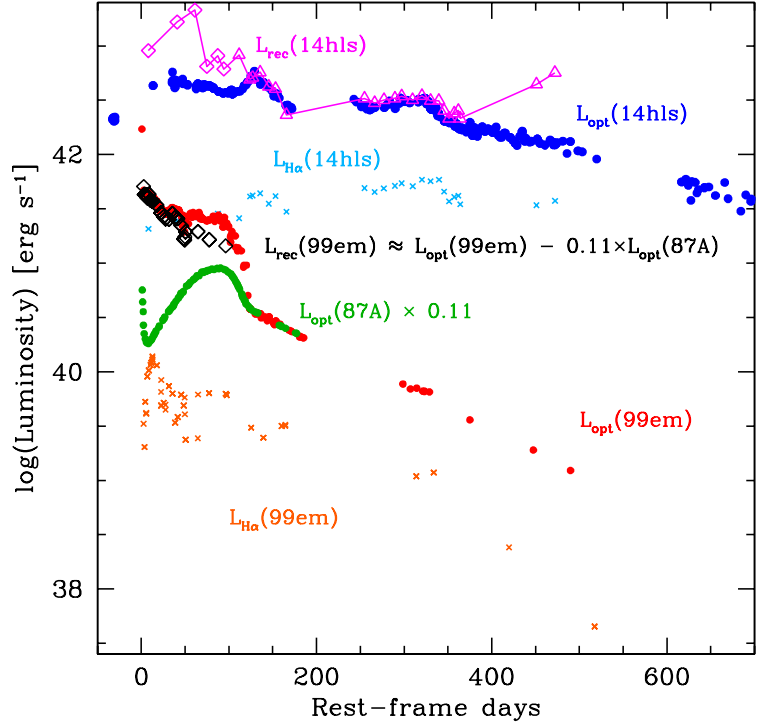}
\caption{Recombination luminosity (magenta open symbols) of iPTF14hls, estimated based on the behaviour of Type~IIP SNe.   The zero time of  iPTF14hls is set to  coincide with the start of the bright phase.
All  optical LCs (filled circles) were computed  in the range $3500-8500$ \AA\ (as described in the text, where we call them "integrated" optical LCs).
The H$\alpha$ luminosities of iPTF14hls and  SN~1999em, estimated from the spectra as detailed in the text,  are reported as light blue and orange crosses, respectively.  The optical  LC of SN~1987A (green circles) was rescaled to match the $^{56}$Co radioactive tail of SN~1999em; the difference between optical luminosities of SN~1999em and normalized SN~1987A is reported as black open diamonds; the recombination luminosity of iPTF14hls, computed from its H$\alpha$ luminosity and from the recombination and H$\alpha$ luminosities of SN~1999em, is represented by magenta diamonds and triangles before and after 110~days, respectively (see text).
}\label{fig:finaltest}
\end{figure*}

We selected the nearby (7.7 Mpc) and well-observed SN~1999em as a prototype Type IIP SN.  First we estimated the pseudo-bolometric luminosities --  or integrated optical luminosities -- of iPTF14hls and SN~1999em  in the range $3500-8500$ \AA\ (reported in Fig.~\ref{fig:finaltest} as blue and red filled circles, respectively).  For iPTF14hls we used  $Bgi$ photometry from \citet{arcavi2017} and \citet{sollerman2019latephase}, after applying a correction for Galactic reddening of $E(B-V) = 0.014$ and using a distance of 145 Mpc; for  SN~1999em we used the $UBVI$ photometry reported in \citealt{hamuy2001,leonard2002,elmhamdi2003,anderson2014,faran2014,galbany2016} (the $R$ filter was avoided as it includes the strong H$\alpha$ emission), dereddened with $E(B-V) = 0.0346$.  From the  dereddened spectra of  iPTF14hls and SN1999em we evaluated the H$\alpha$ emission line net luminosities  for both sources, i.e. we subtracted the absorbed from the emitted flux.  These are reported in  Fig.~\ref{fig:finaltest}  as light blue and orange crosses, respectively. 
In order to compare the light curves of SN1999em and iPTF14hls we set $t=0$ in iPTF14hls to coincide with the beginning of the first bright phase, which we argue is driven by recombination radiation (as is also required by the fact that the peak in luminosity precedes that in the mass-loss rate). 

As it is known that some of the luminosity of a Type~IIP SN during the plateau is due to nuclear decay energy, we removed from the plateau phase  (i.e. up to $\sim110$~days) of the integrated optical LC of SN~1999em the integrated optical LC of SN~1987A  (referring to a similar wavelength range as that adopted for SN~1999em, as it is derived from $UBVRI$ photometry, \citealt{hamuy1988}), which did not have an extended hydrogen envelope and therefore did not display a plateau, but rather a delayed ($\sim 60$~days) luminosity peak caused by the deposition of gamma-rays and positrons and the diffusion of the optical radiation created by their thermalisation.  This subtracted LC is the recombination luminosity of SN~1999em at epochs prior to $\sim110$ days (represented by black open circles in Fig.~\ref{fig:finaltest}), when the plateau dominates the emission.  
We then computed the recombination luminosity in iPTF14hls assuming that it had the same ratio to H$\alpha$ as it did in SN~1999em. The result is shown as open magenta diamonds in Fig.~\ref{fig:finaltest}. This shows that if our assumption is correct, there is more than enough radiation recombination in iPTF14hls to power the early bright phase of the LC. 

After the plateau phase, Type~IIP SN LCs settle on to a linear phase called the radioactive tail, where the energy that is deposited from radioactivity is immediately re-emitted and diffuses out as optical radiation without delay. Therefore, we computed the ratio of the integrated optical luminosity to the
H$\alpha$ luminosity  in SN~1999em after 110~days and multiplied it to the H$\alpha$ luminosity in iPTF14hls,  to check whether similar conditions apply in the two events, as the spectral similarity suggests: the increasing prevalence of H$\alpha$ emission suggests an increasingly transparent flow, where the underlying luminosity (deposited radioactive energy in the Type~IIP SN and stellar luminosity in iPTF14hls) is reflected immediately in the luminosity of the outflow and diffusion times are negligible. The estimated recombination luminosity of iPTF14hls past 110~days (open magenta triangles in Fig.~\ref{fig:finaltest})  shows  that this approximation is also reasonably good. 
Therefore, iPTF14hls in the early phase can be thought of as obeying radiation laws of a Type~IIP SN without being a SN: it is a massive outflow which progressively decreases in mass flow as the driving stellar luminosity fades.

\subsubsection{Late phase luminosity}
At very late phases, when it faded after the 2-year bright phase, iPTF14hls showed observational signatures of circumstellar interaction \citep{andrew2018latephase,sollerman2019latephase}.
Circumstellar interaction would naturally be expected after the extraordinary mass loss from the progenitor because the outflow can crash into the pre-existing circumstellar matter \citep[e.g.,][]{woosley2007}.
The possible brightening of the progenitor of iPTF14hls in 1954 \citep{arcavi2017} may represent a previous pulsational event and may be responsible for the formation of an outer dense shell of circumstellar matter. Thus, the source of luminosity of
iPTF14hls in the late phase could be interaction converting outflow kinetic energy
into radiation. The effect of the collision was not significant during the bright
phase, which was powered by a very high stellar luminosity reprocessed
via hydrogen ionization and the ensuing recombination, as for Type~IIP SNe.
The interaction became apparent after iPTF14hls faded.

\subsection{Energy source}
We argue that iPTF14hls is a hyper-wind rather than a mass eruption. Given the extremely large luminosity far beyond the Eddington luminosity, the wind is probably driven by radiation as suggested for \ec\ and other luminous blue variables \citep[e.g.,][]{smith2006owocki}. The total estimated kinetic energy of the outflow is estimated to be $10^{52}~\mathrm{erg}$ and it is not clear how the progenitor gained such a huge energy to drive the hyper-wind.

Pulsational pair-instability is one possibility \citep{woosley2007,woosley2017,marchant2018,vigna-gomez2019}. Pulsational pair-instability SNe (PPISNe) have already been related to iPTF14hls in previous studies \citep[e.g.,][]{woosley2018}. PPISNe from progenitors having zero-age main-sequence mass of around 110~\Msun\ have pulsation duration of several years \citep{woosley2017} matching the duration of the continuum-driven wind in iPTF14hls as well as the estimated progenitor mass (Section~\ref{sec:greateruption}). A caveat is that the pulsational pair-instability may just lead to a violent mass ejection rather than a continuous mass outflow. Also, the PPISN mass outflows are predicted to have kinetic energy up to $3\times 10^{51}~\mathrm{erg}$ \citep{woosley2017}, which is about a factor 3 smaller than the energy deduced for iPTF14hls ($10^{52}~\mathrm{erg}$). Given the uncertainties in modelling the pulsation \citep[e.g.,][]{takahashi2016}, however, it is possible that pulsational instabilities may achieve such a high energy. Such an energetic pulsation of a massive star is predicted to be followed by the formation of a black-hole. This may indeed have been the fate of the progenitor of iPTF14hls. Later observations may reveal whether any source is present at the location of iPTF14hls.

It has been suggested that the Great Eruption of \ec\ is a result of a stellar merger in a close binary system triggered by a third massive star \citep[e.g.,][]{PortegiesZwart2016}. In such a case, a strong continuum-driven wind is suggested to occur following the release of tidal energy dissipation prior to the merger. The variability in the mass-loss rate may be related to the rotational period of the eccentric merging binary system. The total energy available to blow the wind depends on the binary mass and separation and it can, in principle, reach $10^{52}~\mathrm{erg}$. The huge difference in the luminosity between iPTF14hls and \ec\ could be due to the much closer initial binary separation in iPTF14hls. The timescale of the wind could also be affected by the initial binary configuration and the properties of the third massive star.

Finally, the instability at the surface of massive red supergiants (RSGs) that is triggered when the luminosity-to-mass ratio is high is another possible mechanism to initiate a strong wind \citep{heger1997superwind,yoon2010superwind}. \citet{moriya2015superwind} investigated the surface instability of a 150~\Msun\ RSG and found that mass-loss enhancement triggered by the surface instability could last for several years. However, the estimated mass-loss rates from the pulsation were up to $\sim 0.1~\Msunpyr$ and did not go beyond $1~\Msunpyr$ as estimated for iPTF14hls. However, how the pulsation drives the wind is uncertain.

\section{Conclusions}
We have proposed that iPTF14hls is a continuous outflow like a stellar wind rather than a mass ejection like a SN, and unveiled its mass-loss rate history. The slow change in its spectroscopic properties over 2 years is naturally explained by such a continuous wind. The mass-loss rates exceed a few~\Msunpyr\ during the bright phase of iPTF14hls and temporarily become as high as 10~\Msunpyr. This hyper-wind is a super-Eddington continuum-driven wind in which the hydrogen recombination likely powers the early bright phase of iPTF14hls. We have shown that \ec\ during the Great Eruption share similar properties to iPTF14hls including their mass-loss rates and LC shapes. The two massive stars may have similar mass (around 100~\Msun) as well. The exact luminosity source of iPTF14hls is not clear, but PPISNe or massive close binary mergers may be related as suggested for the Great Eruption of \ec.

\section*{Acknowledgements}
PAM and EP are grateful for kind hospitality at NAOJ during completion of this work. This work was supported by NAOJ Research Coordination Committee, NINS, Grant Number 19FS-0506 and 19FS-0507. TJM is supported by the Grants-in-Aid for Scientific Research of the Japan Society for the Promotion of Science (JP17H02864, JP18K13585).
We made use of WISeREP (\url{https://wiserep.weizmann.ac.il}, \citealt{yaron2012wiserep}) and the Open Supernova Catalog (\url{https://sne.space/}, \citealt{guillochon2017osn}).




\bibliographystyle{mnras}
\bibliography{mnras} 

\begin{thebibliography}{}
\makeatletter
\relax
\def\mn@urlcharsother{\let\do\@makeother \do\$\do\&\do\#\do\^\do\_\do\%\do\~}
\def\mn@doi{\begingroup\mn@urlcharsother \@ifnextchar [ {\mn@doi@}
  {\mn@doi@[]}}
\def\mn@doi@[#1]#2{\def\@tempa{#1}\ifx\@tempa\@empty \href
  {http://dx.doi.org/#2} {doi:#2}\else \href {http://dx.doi.org/#2} {#1}\fi
  \endgroup}
\def\mn@eprint#1#2{\mn@eprint@#1:#2::\@nil}
\def\mn@eprint@arXiv#1{\href {http://arxiv.org/abs/#1} {{\tt arXiv:#1}}}
\def\mn@eprint@dblp#1{\href {http://dblp.uni-trier.de/rec/bibtex/#1.xml}
  {dblp:#1}}
\def\mn@eprint@#1:#2:#3:#4\@nil{\def\@tempa {#1}\def\@tempb {#2}\def\@tempc
  {#3}\ifx \@tempc \@empty \let \@tempc \@tempb \let \@tempb \@tempa \fi \ifx
  \@tempb \@empty \def\@tempb {arXiv}\fi \@ifundefined
  {mn@eprint@\@tempb}{\@tempb:\@tempc}{\expandafter \expandafter \csname
  mn@eprint@\@tempb\endcsname \expandafter{\@tempc}}}

\bibitem[\protect\citeauthoryear{{Anderson} et~al.,}{{Anderson}
  et~al.}{2014}]{anderson2014}
{Anderson} J.~P.,  et~al., 2014, \mn@doi [\apj] {10.1088/0004-637X/786/1/67},
  \href {https://ui.adsabs.harvard.edu/abs/2014ApJ...786...67A} {786, 67}

\bibitem[\protect\citeauthoryear{{Andrews} \& {Smith}}{{Andrews} \&
  {Smith}}{2018}]{andrew2018latephase}
{Andrews} J.~E.,  {Smith} N.,  2018, \mn@doi [\mnras] {10.1093/mnras/sty584},
  \href {https://ui.adsabs.harvard.edu/abs/2018MNRAS.477...74A} {477, 74}

\bibitem[\protect\citeauthoryear{{Arcavi} et~al.,}{{Arcavi}
  et~al.}{2017}]{arcavi2017}
{Arcavi} I.,  et~al., 2017, \mn@doi [Nature] {10.1038/nature24030}, \href
  {https://ui.adsabs.harvard.edu/abs/2017Natur.551..210A} {551, 210}

\bibitem[\protect\citeauthoryear{{Bennett}}{{Bennett}}{2010}]{bennett2010rsgwind}
{Bennett} P.~D.,  2010, in {Leitherer} C.,  {Bennett} P.~D.,  {Morris} P.~W.,
  {Van Loon} J.~T.,  eds,  Astronomical Society of the Pacific Conference
  Series Vol. 425, Hot and Cool: Bridging Gaps in Massive Star Evolution.
  p.~181 (\mn@eprint {arXiv} {1004.1853})

\bibitem[\protect\citeauthoryear{{Bersten}, {Benvenuto}  \& {Hamuy}}{{Bersten}
  et~al.}{2011}]{bersten2011iip}
{Bersten} M.~C.,  {Benvenuto} O.,   {Hamuy} M.,  2011, \mn@doi [\apj]
  {10.1088/0004-637X/729/1/61}, \href
  {https://ui.adsabs.harvard.edu/abs/2011ApJ...729...61B} {729, 61}

\bibitem[\protect\citeauthoryear{{Chugai}}{{Chugai}}{2018}]{chugai2018}
{Chugai} N.~N.,  2018, \mn@doi [Astronomy Letters] {10.1134/S1063773718060026},
  \href {https://ui.adsabs.harvard.edu/abs/2018AstL...44..370C} {44, 370}

\bibitem[\protect\citeauthoryear{{Davidson}}{{Davidson}}{1987}]{davidson1987}
{Davidson} K.,  1987, \mn@doi [\apj] {10.1086/165324}, \href
  {https://ui.adsabs.harvard.edu/abs/1987ApJ...317..760D} {317, 760}

\bibitem[\protect\citeauthoryear{{Davidson} \& {Humphreys}}{{Davidson} \&
  {Humphreys}}{1997}]{davidson1997ecar}
{Davidson} K.,  {Humphreys} R.~M.,  1997, \mn@doi [\araa]
  {10.1146/annurev.astro.35.1.1}, \href
  {https://ui.adsabs.harvard.edu/abs/1997ARA&A..35....1D} {35, 1}

\bibitem[\protect\citeauthoryear{{Dessart}}{{Dessart}}{2018}]{dessart2018mag}
{Dessart} L.,  2018, \mn@doi [\aap] {10.1051/0004-6361/201732402}, \href
  {https://ui.adsabs.harvard.edu/abs/2018A&A...610L..10D} {610, L10}

\bibitem[\protect\citeauthoryear{{Eastman}, {Schmidt}  \& {Kirshner}}{{Eastman}
  et~al.}{1996}]{eastman1996}
{Eastman} R.~G.,  {Schmidt} B.~P.,   {Kirshner} R.,  1996, \mn@doi [\apj]
  {10.1086/177563}, \href
  {https://ui.adsabs.harvard.edu/abs/1996ApJ...466..911E} {466, 911}

\bibitem[\protect\citeauthoryear{{Ekstr{\"o}m} et~al.,}{{Ekstr{\"o}m}
  et~al.}{2012}]{ekstrom2012}
{Ekstr{\"o}m} S.,  et~al., 2012, \mn@doi [\aap] {10.1051/0004-6361/201117751},
  \href {https://ui.adsabs.harvard.edu/abs/2012A&A...537A.146E} {537, A146}

\bibitem[\protect\citeauthoryear{{Elmhamdi} et~al.,}{{Elmhamdi}
  et~al.}{2003}]{elmhamdi2003}
{Elmhamdi} A.,  et~al., 2003, \mn@doi [\mnras]
  {10.1046/j.1365-8711.2003.06150.x}, \href
  {https://ui.adsabs.harvard.edu/abs/2003MNRAS.338..939E} {338, 939}

\bibitem[\protect\citeauthoryear{{Faran} et~al.,}{{Faran}
  et~al.}{2014}]{faran2014}
{Faran} T.,  et~al., 2014, \mn@doi [\mnras] {10.1093/mnras/stu955}, \href
  {https://ui.adsabs.harvard.edu/abs/2014MNRAS.442..844F} {442, 844}

\bibitem[\protect\citeauthoryear{{Fern{\'a}ndez-Laj{\'u}s}
  et~al.,}{{Fern{\'a}ndez-Laj{\'u}s}
  et~al.}{2009}]{fernandez-lajus2009etacarlightcurve}
{Fern{\'a}ndez-Laj{\'u}s} E.,  et~al., 2009, \mn@doi [\aap]
  {10.1051/0004-6361:200810700}, \href
  {https://ui.adsabs.harvard.edu/abs/2009A&A...493.1093F} {493, 1093}

\bibitem[\protect\citeauthoryear{{Filippenko}}{{Filippenko}}{1997}]{filippenko1997}
{Filippenko} A.~V.,  1997, \mn@doi [\araa] {10.1146/annurev.astro.35.1.309},
  \href {https://ui.adsabs.harvard.edu/abs/1997ARA&A..35..309F} {35, 309}

\bibitem[\protect\citeauthoryear{{Galbany} et~al.,}{{Galbany}
  et~al.}{2016}]{galbany2016}
{Galbany} L.,  et~al., 2016, \mn@doi [\aj] {10.3847/0004-6256/151/2/33}, \href
  {https://ui.adsabs.harvard.edu/abs/2016AJ....151...33G} {151, 33}

\bibitem[\protect\citeauthoryear{{Georgy} et~al.,}{{Georgy}
  et~al.}{2013}]{georgy2013}
{Georgy} C.,  et~al., 2013, \mn@doi [\aap] {10.1051/0004-6361/201322178}, \href
  {https://ui.adsabs.harvard.edu/abs/2013A&A...558A.103G} {558, A103}

\bibitem[\protect\citeauthoryear{{Gofman} \& {Soker}}{{Gofman} \&
  {Soker}}{2019}]{gofman2019binary}
{Gofman} R.~A.,  {Soker} N.,  2019, arXiv e-prints, \href
  {https://ui.adsabs.harvard.edu/abs/2019arXiv190505573G} {p. arXiv:1905.05573}

\bibitem[\protect\citeauthoryear{{Guillochon}, {Parrent}, {Kelley}  \&
  {Margutti}}{{Guillochon} et~al.}{2017}]{guillochon2017osn}
{Guillochon} J.,  {Parrent} J.,  {Kelley} L.~Z.,   {Margutti} R.,  2017,
  \mn@doi [\apj] {10.3847/1538-4357/835/1/64}, \href
  {https://ui.adsabs.harvard.edu/abs/2017ApJ...835...64G} {835, 64}

\bibitem[\protect\citeauthoryear{{Hamuy}, {Suntzeff}, {Gonzalez}  \&
  {Martin}}{{Hamuy} et~al.}{1988}]{hamuy1988}
{Hamuy} M.,  {Suntzeff} N.~B.,  {Gonzalez} R.,   {Martin} G.,  1988, \mn@doi
  [\aj] {10.1086/114613}, \href
  {https://ui.adsabs.harvard.edu/abs/1988AJ.....95...63H} {95, 63}

\bibitem[\protect\citeauthoryear{{Hamuy} et~al.,}{{Hamuy}
  et~al.}{2001}]{hamuy2001}
{Hamuy} M.,  et~al., 2001, \mn@doi [\apj] {10.1086/322450}, \href
  {https://ui.adsabs.harvard.edu/abs/2001ApJ...558..615H} {558, 615}

\bibitem[\protect\citeauthoryear{{Heger}, {Jeannin}, {Langer}  \&
  {Baraffe}}{{Heger} et~al.}{1997}]{heger1997superwind}
{Heger} A.,  {Jeannin} L.,  {Langer} N.,   {Baraffe} I.,  1997, \aap, \href
  {https://ui.adsabs.harvard.edu/abs/1997A&A...327..224H} {327, 224}

\bibitem[\protect\citeauthoryear{{Kasen} \& {Woosley}}{{Kasen} \&
  {Woosley}}{2009}]{kasen2009iip}
{Kasen} D.,  {Woosley} S.~E.,  2009, \mn@doi [\apj]
  {10.1088/0004-637X/703/2/2205}, \href
  {https://ui.adsabs.harvard.edu/abs/2009ApJ...703.2205K} {703, 2205}

\bibitem[\protect\citeauthoryear{{Leonard} et~al.,}{{Leonard}
  et~al.}{2002}]{leonard2002}
{Leonard} D.~C.,  et~al., 2002, \mn@doi [\pasp] {10.1086/324785}, \href
  {https://ui.adsabs.harvard.edu/abs/2002PASP..114...35L} {114, 35}

\bibitem[\protect\citeauthoryear{{Liu}, {Song}, {Yi}, {Gu}  \& {Wang}}{{Liu}
  et~al.}{2019}]{liu2019fallback}
{Liu} T.,  {Song} C.-Y.,  {Yi} T.,  {Gu} W.-M.,   {Wang} X.-F.,  2019, \mn@doi
  [Journal of High Energy Astrophysics] {10.1016/j.jheap.2019.02.001}, \href
  {https://ui.adsabs.harvard.edu/abs/2019JHEAp..22....5L} {22, 5}

\bibitem[\protect\citeauthoryear{{Marchant}, {Renzo}, {Farmer}, {Pappas},
  {Taam}, {de Mink}  \& {Kalogera}}{{Marchant} et~al.}{2018}]{marchant2018}
{Marchant} P.,  {Renzo} M.,  {Farmer} R.,  {Pappas} K. M.~W.,  {Taam} R.~E.,
  {de Mink} S.,   {Kalogera} V.,  2018, arXiv e-prints, \href
  {https://ui.adsabs.harvard.edu/abs/2018arXiv181013412M} {p. arXiv:1810.13412}

\bibitem[\protect\citeauthoryear{{Mazzali}, {Deng}, {Maeda}, {Nomoto},
  {Filippenko}  \& {Matheson}}{{Mazzali} et~al.}{2004}]{mazzali04eo}
{Mazzali} P.~A.,  {Deng} J.,  {Maeda} K.,  {Nomoto} K.,  {Filippenko} A.~V.,
  {Matheson} T.,  2004, \mn@doi [\apj] {10.1086/423888}, \href
  {https://ui.adsabs.harvard.edu/abs/2004ApJ...614..858M} {614, 858}

\bibitem[\protect\citeauthoryear{{Moriya} \& {Langer}}{{Moriya} \&
  {Langer}}{2015}]{moriya2015superwind}
{Moriya} T.~J.,  {Langer} N.,  2015, \mn@doi [\aap]
  {10.1051/0004-6361/201424957}, \href
  {https://ui.adsabs.harvard.edu/abs/2015A&A...573A..18M} {573, A18}

\bibitem[\protect\citeauthoryear{{Morris} et~al.,}{{Morris}
  et~al.}{1999}]{morris1999}
{Morris} P.~W.,  et~al., 1999, \mn@doi [Nature] {10.1038/990048}, \href
  {https://ui.adsabs.harvard.edu/abs/1999Natur.402..502M} {402, 502}

\bibitem[\protect\citeauthoryear{{Owocki} \& {Shaviv}}{{Owocki} \&
  {Shaviv}}{2016}]{owocki2016eccdw}
{Owocki} S.~P.,  {Shaviv} N.~J.,  2016, \mn@doi [\mnras]
  {10.1093/mnras/stw1642}, \href
  {https://ui.adsabs.harvard.edu/abs/2016MNRAS.462..345O} {462, 345}

\bibitem[\protect\citeauthoryear{{Pejcha} \& {Prieto}}{{Pejcha} \&
  {Prieto}}{2015}]{pejcha2015}
{Pejcha} O.,  {Prieto} J.~L.,  2015, \mn@doi [\apj]
  {10.1088/0004-637X/806/2/225}, \href
  {https://ui.adsabs.harvard.edu/abs/2015ApJ...806..225P} {806, 225}

\bibitem[\protect\citeauthoryear{{Portegies Zwart} \& {van den
  Heuvel}}{{Portegies Zwart} \& {van den Heuvel}}{2016}]{PortegiesZwart2016}
{Portegies Zwart} S.~F.,  {van den Heuvel} E.~P.~J.,  2016, \mn@doi [\mnras]
  {10.1093/mnras/stv2787}, \href
  {https://ui.adsabs.harvard.edu/abs/2016MNRAS.456.3401P} {456, 3401}

\bibitem[\protect\citeauthoryear{{Prieto} et~al.,}{{Prieto}
  et~al.}{2014}]{prieto2014ececho}
{Prieto} J.~L.,  et~al., 2014, \mn@doi [\apj] {10.1088/2041-8205/787/1/L8},
  \href {https://ui.adsabs.harvard.edu/abs/2014ApJ...787L...8P} {787, L8}

\bibitem[\protect\citeauthoryear{{Rest} et~al.,}{{Rest}
  et~al.}{2012}]{rest2012echo}
{Rest} A.,  et~al., 2012, \mn@doi [Nature] {10.1038/nature10775}, \href
  {https://ui.adsabs.harvard.edu/abs/2012Natur.482..375R} {482, 375}

\bibitem[\protect\citeauthoryear{{Smith}}{{Smith}}{2006}]{smith2006homu}
{Smith} N.,  2006, \mn@doi [\apj] {10.1086/503766}, \href
  {https://ui.adsabs.harvard.edu/abs/2006ApJ...644.1151S} {644, 1151}

\bibitem[\protect\citeauthoryear{{Smith}}{{Smith}}{2013}]{smith2013ec}
{Smith} N.,  2013, \mn@doi [\mnras] {10.1093/mnras/sts508}, \href
  {https://ui.adsabs.harvard.edu/abs/2013MNRAS.429.2366S} {429, 2366}

\bibitem[\protect\citeauthoryear{{Smith} \& {Owocki}}{{Smith} \&
  {Owocki}}{2006}]{smith2006owocki}
{Smith} N.,  {Owocki} S.~P.,  2006, \mn@doi [\apjl] {10.1086/506523}, \href
  {https://ui.adsabs.harvard.edu/abs/2006ApJ...645L..45S} {645, L45}

\bibitem[\protect\citeauthoryear{{Smith} et~al.,}{{Smith}
  et~al.}{2018}]{smith2018highvec}
{Smith} N.,  et~al., 2018, \mn@doi [\mnras] {10.1093/mnras/sty1479}, \href
  {https://ui.adsabs.harvard.edu/abs/2018MNRAS.480.1457S} {480, 1457}

\bibitem[\protect\citeauthoryear{{Soker} \& {Gilkis}}{{Soker} \&
  {Gilkis}}{2018}]{soker2018comenvjet}
{Soker} N.,  {Gilkis} A.,  2018, \mn@doi [\mnras] {10.1093/mnras/stx3287},
  \href {https://ui.adsabs.harvard.edu/abs/2018MNRAS.475.1198S} {475, 1198}

\bibitem[\protect\citeauthoryear{{Sollerman} et~al.,}{{Sollerman}
  et~al.}{2019}]{sollerman2019latephase}
{Sollerman} J.,  et~al., 2019, \mn@doi [\aap] {10.1051/0004-6361/201833689},
  \href {https://ui.adsabs.harvard.edu/abs/2019A&A...621A..30S} {621, A30}

\bibitem[\protect\citeauthoryear{{Sz{\'e}csi}, {Langer}, {Yoon}, {Sanyal}, {de
  Mink}, {Evans}  \& {Dermine}}{{Sz{\'e}csi} et~al.}{2015}]{szecsi2015lowmetal}
{Sz{\'e}csi} D.,  {Langer} N.,  {Yoon} S.-C.,  {Sanyal} D.,  {de Mink} S.,
  {Evans} C.~J.,   {Dermine} T.,  2015, \mn@doi [\aap]
  {10.1051/0004-6361/201526617}, \href
  {https://ui.adsabs.harvard.edu/abs/2015A&A...581A..15S} {581, A15}

\bibitem[\protect\citeauthoryear{{Takahashi}, {Yoshida}, {Umeda}, {Sumiyoshi}
  \& {Yamada}}{{Takahashi} et~al.}{2016}]{takahashi2016}
{Takahashi} K.,  {Yoshida} T.,  {Umeda} H.,  {Sumiyoshi} K.,   {Yamada} S.,
  2016, \mn@doi [\mnras] {10.1093/mnras/stv2649}, \href
  {https://ui.adsabs.harvard.edu/abs/2016MNRAS.456.1320T} {456, 1320}

\bibitem[\protect\citeauthoryear{{Vigna-G{\'o}mez}, {Justham}, {Mandel}, {de
  Mink}  \& {Podsiadlowski}}{{Vigna-G{\'o}mez} et~al.}{2019}]{vigna-gomez2019}
{Vigna-G{\'o}mez} A.,  {Justham} S.,  {Mandel} I.,  {de Mink} S.~E.,
  {Podsiadlowski} P.,  2019, \mn@doi [\apjl] {10.3847/2041-8213/ab1bdf}, \href
  {https://ui.adsabs.harvard.edu/abs/2019ApJ...876L..29V} {876, L29}

\bibitem[\protect\citeauthoryear{{Wang} et~al.,}{{Wang}
  et~al.}{2018}]{wang2018fallback}
{Wang} L.~J.,  et~al., 2018, \mn@doi [\apj] {10.3847/1538-4357/aadba4}, \href
  {https://ui.adsabs.harvard.edu/abs/2018ApJ...865...95W} {865, 95}

\bibitem[\protect\citeauthoryear{{Woosley}}{{Woosley}}{2017}]{woosley2017}
{Woosley} S.~E.,  2017, \mn@doi [\apj] {10.3847/1538-4357/836/2/244}, \href
  {https://ui.adsabs.harvard.edu/abs/2017ApJ...836..244W} {836, 244}

\bibitem[\protect\citeauthoryear{{Woosley}}{{Woosley}}{2018}]{woosley2018}
{Woosley} S.~E.,  2018, \mn@doi [\apj] {10.3847/1538-4357/aad044}, \href
  {https://ui.adsabs.harvard.edu/abs/2018ApJ...863..105W} {863, 105}

\bibitem[\protect\citeauthoryear{{Woosley}, {Blinnikov}  \& {Heger}}{{Woosley}
  et~al.}{2007}]{woosley2007}
{Woosley} S.~E.,  {Blinnikov} S.,   {Heger} A.,  2007, \mn@doi [Nature]
  {10.1038/nature06333}, \href
  {https://ui.adsabs.harvard.edu/abs/2007Natur.450..390W} {450, 390}

\bibitem[\protect\citeauthoryear{{Yaron} \& {Gal-Yam}}{{Yaron} \&
  {Gal-Yam}}{2012}]{yaron2012wiserep}
{Yaron} O.,  {Gal-Yam} A.,  2012, \mn@doi [\pasp] {10.1086/666656}, \href
  {https://ui.adsabs.harvard.edu/abs/2012PASP..124..668Y} {124, 668}

\bibitem[\protect\citeauthoryear{{Yoon} \& {Cantiello}}{{Yoon} \&
  {Cantiello}}{2010}]{yoon2010superwind}
{Yoon} S.-C.,  {Cantiello} M.,  2010, \mn@doi [\apjl]
  {10.1088/2041-8205/717/1/L62}, \href
  {https://ui.adsabs.harvard.edu/abs/2010ApJ...717L..62Y} {717, L62}

\bibitem[\protect\citeauthoryear{{Yoon}, {Dierks}  \& {Langer}}{{Yoon}
  et~al.}{2012}]{yoon2012popiii}
{Yoon} S.~C.,  {Dierks} A.,   {Langer} N.,  2012, \mn@doi [\aap]
  {10.1051/0004-6361/201117769}, \href
  {https://ui.adsabs.harvard.edu/abs/2012A&A...542A.113Y} {542, A113}

\bibitem[\protect\citeauthoryear{{van Marle}, {Owocki}  \& {Shaviv}}{{van
  Marle} et~al.}{2008}]{vanmarle2008supewind}
{van Marle} A.~J.,  {Owocki} S.~P.,   {Shaviv} N.~J.,  2008, \mn@doi [\mnras]
  {10.1111/j.1365-2966.2008.13648.x}, \href
  {https://ui.adsabs.harvard.edu/abs/2008MNRAS.389.1353V} {389, 1353}

\makeatother
\end{thebibliography}




\appendix

\section{Results of the blackbody fitting}\label{app:bbfit}
\begin{figure*}
\includegraphics[width=0.66\columnwidth]{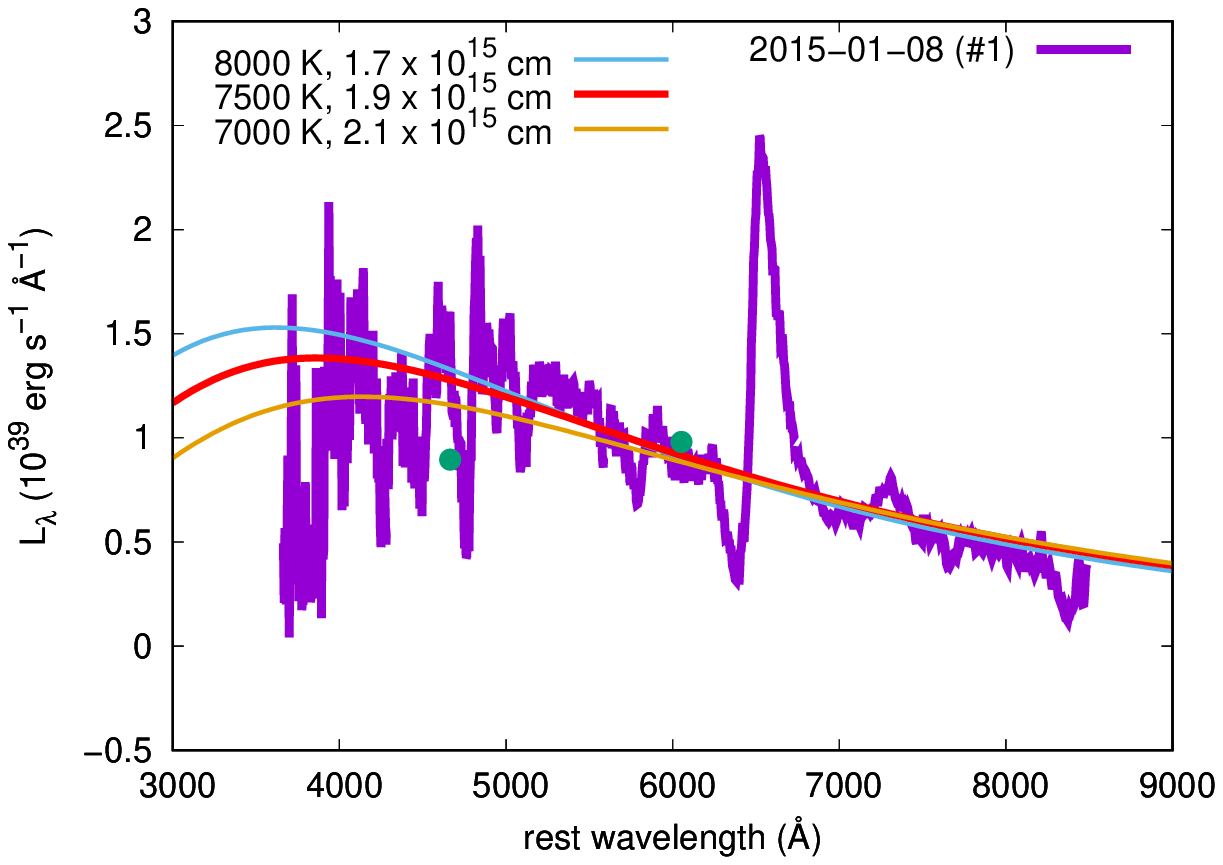} 
\includegraphics[width=0.66\columnwidth]{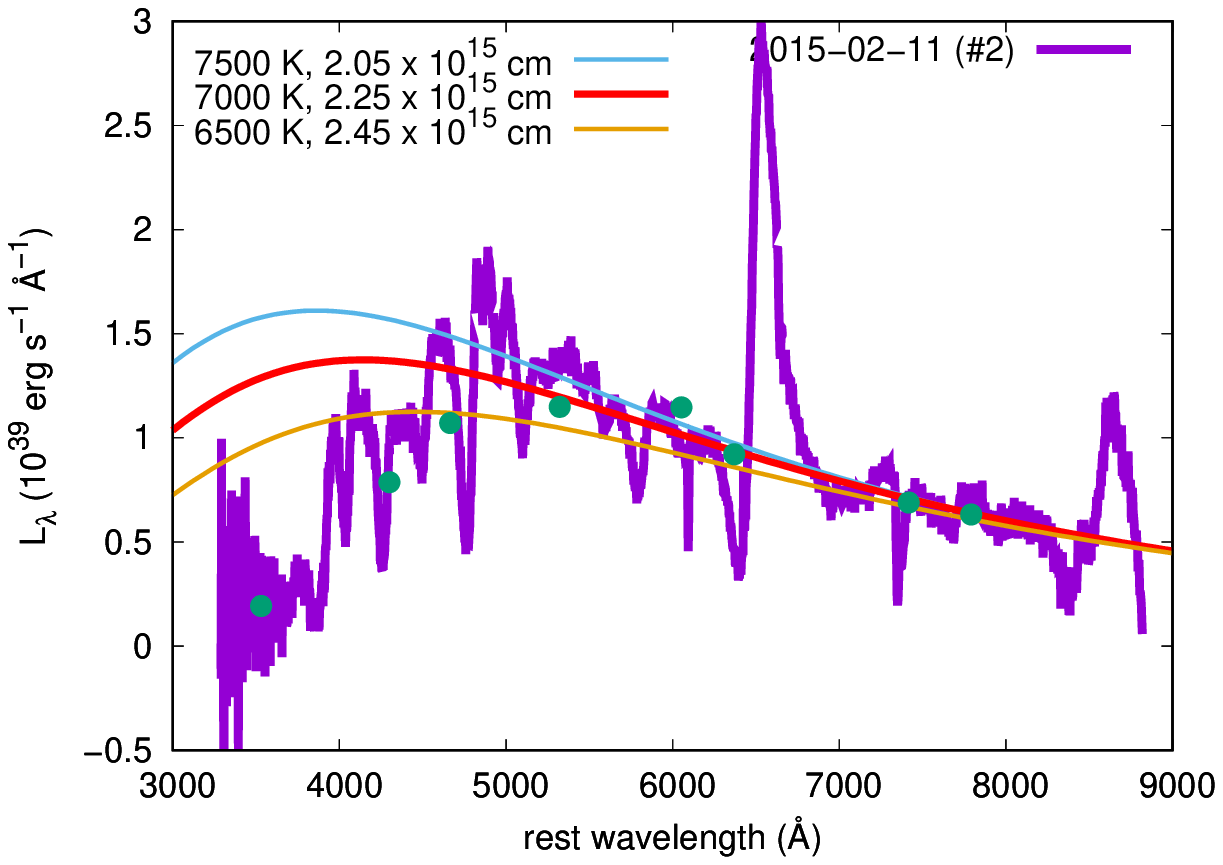}
\includegraphics[width=0.66\columnwidth]{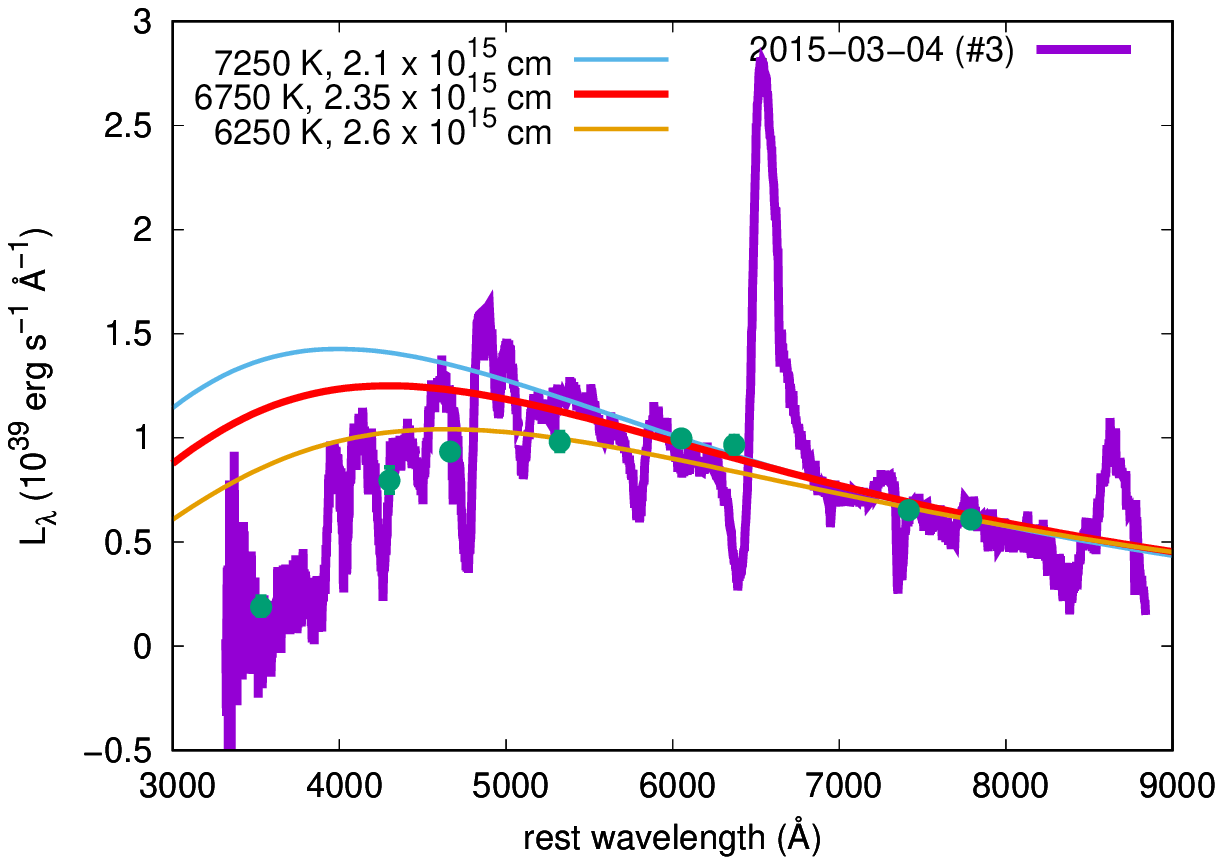}
\includegraphics[width=0.66\columnwidth]{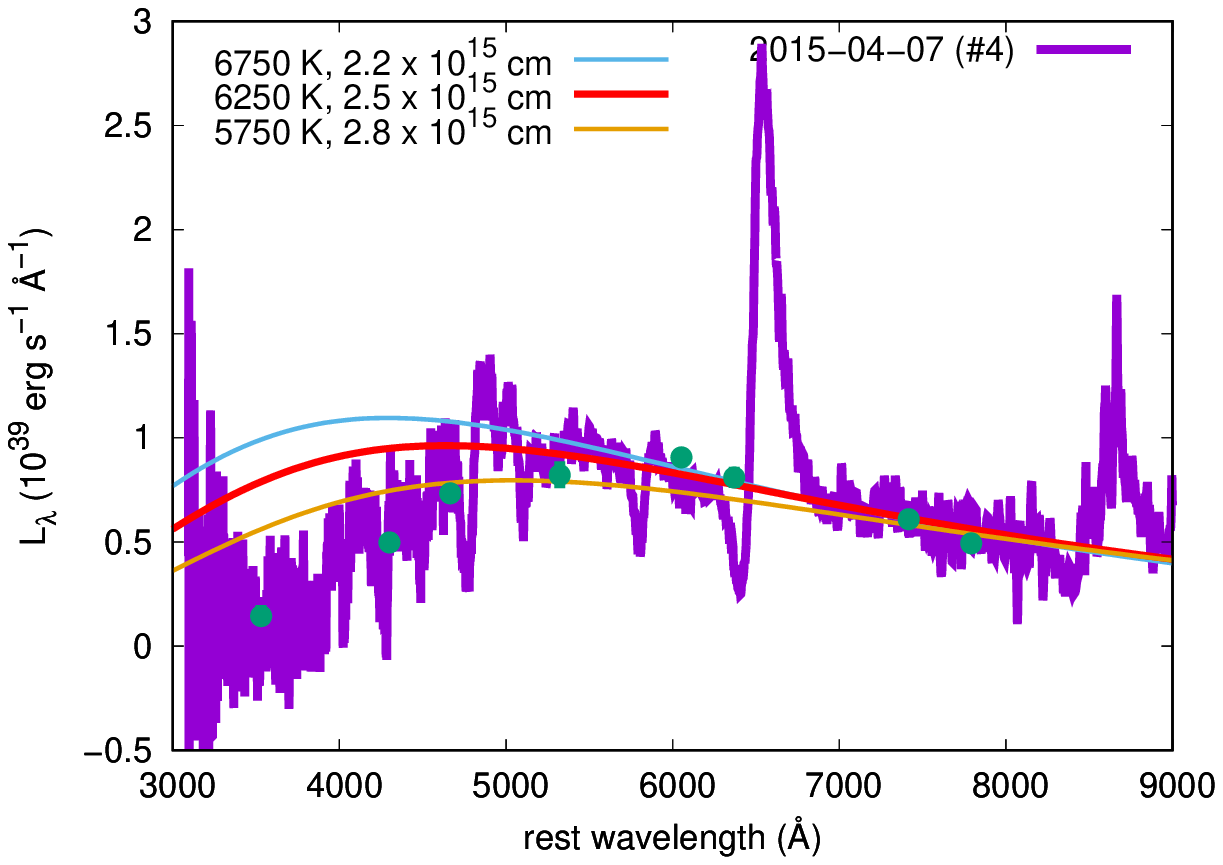} 
\includegraphics[width=0.66\columnwidth]{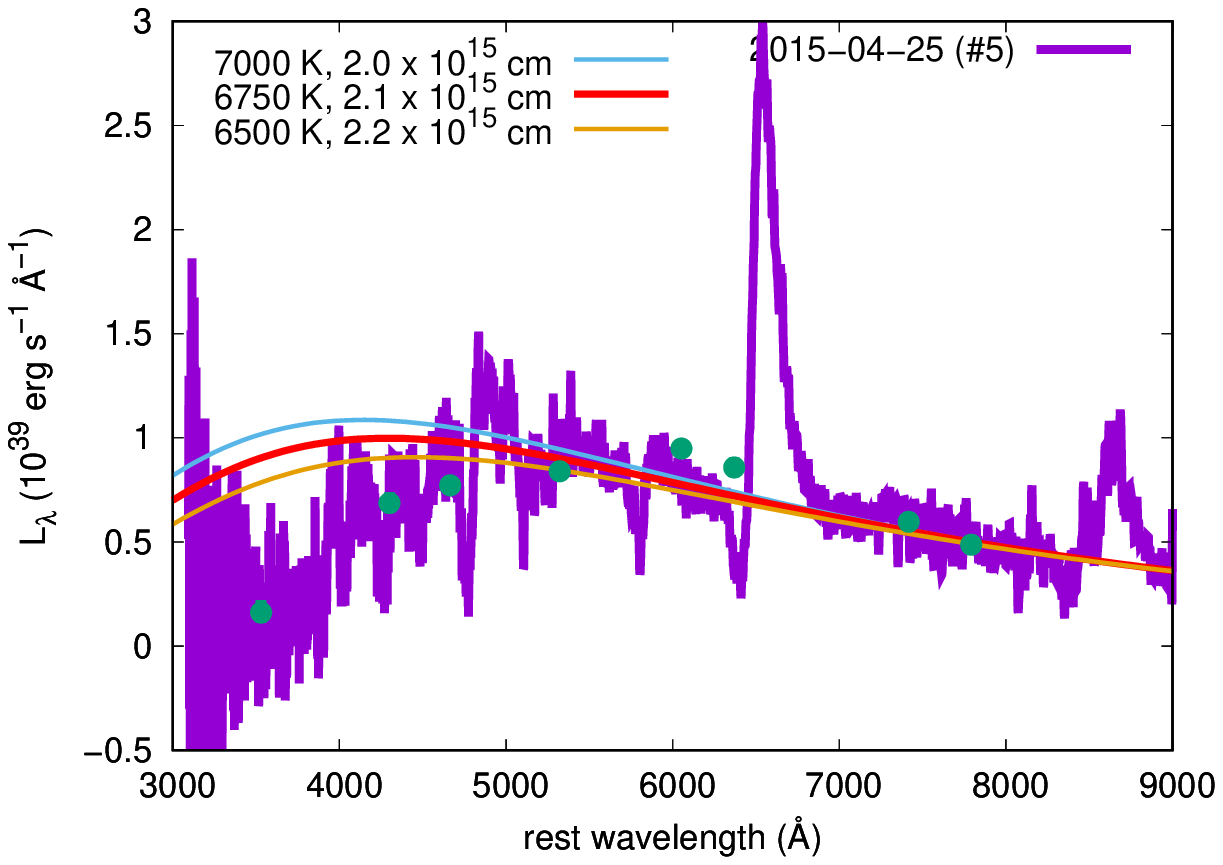} 
\includegraphics[width=0.66\columnwidth]{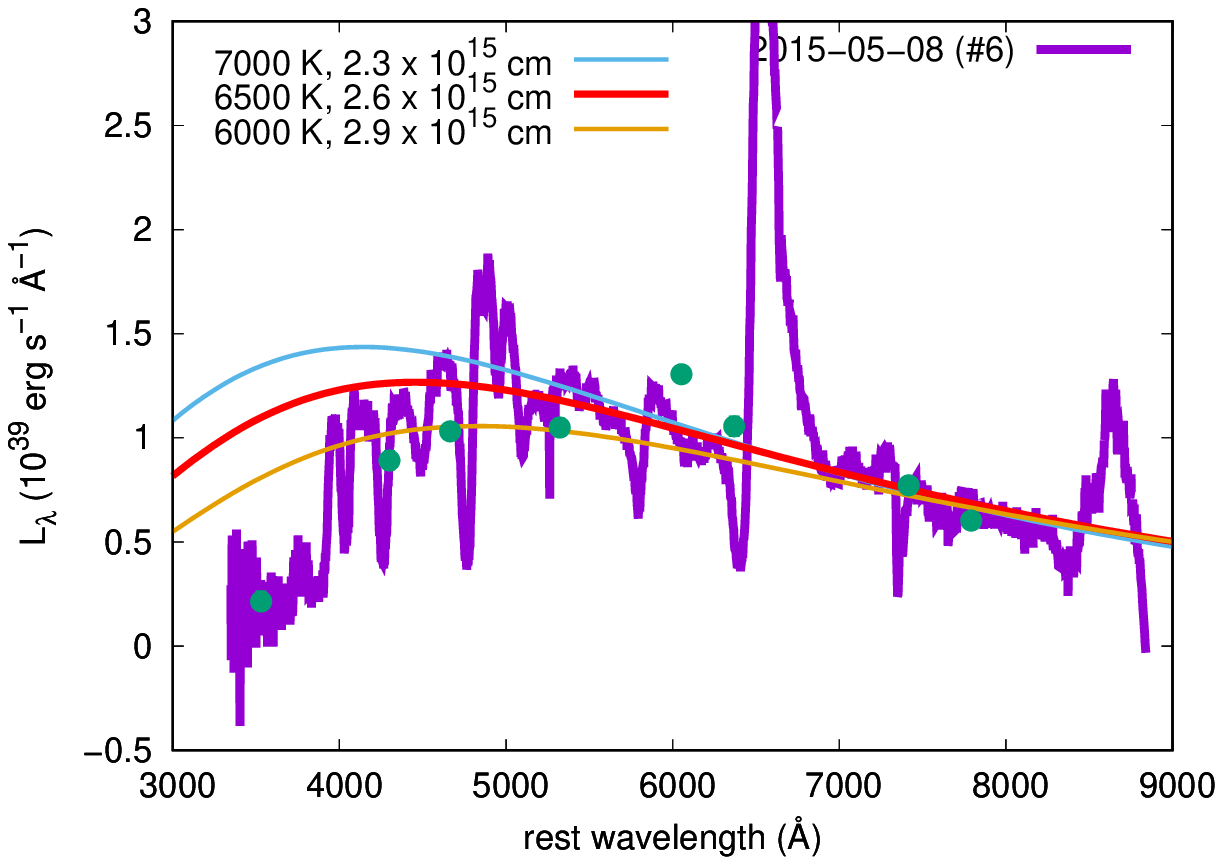} 
\includegraphics[width=0.66\columnwidth]{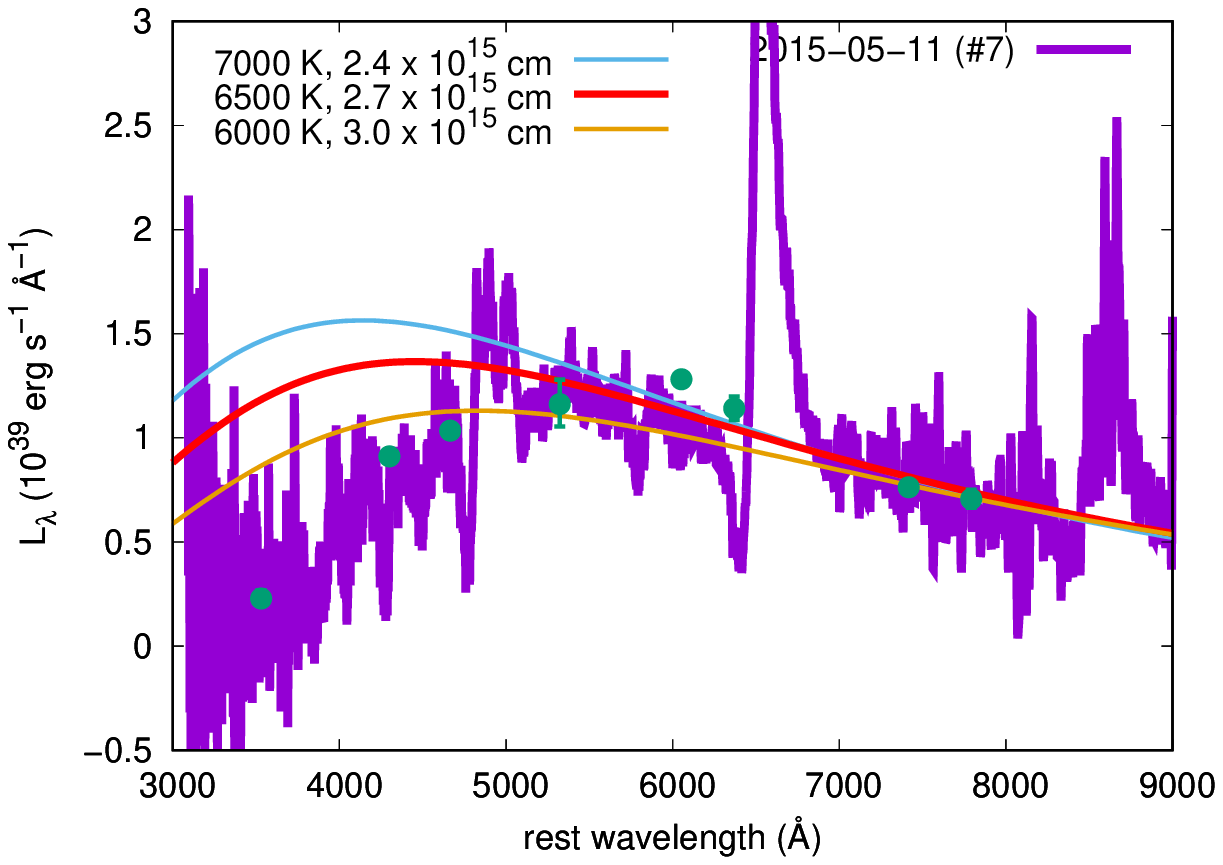} 
\includegraphics[width=0.66\columnwidth]{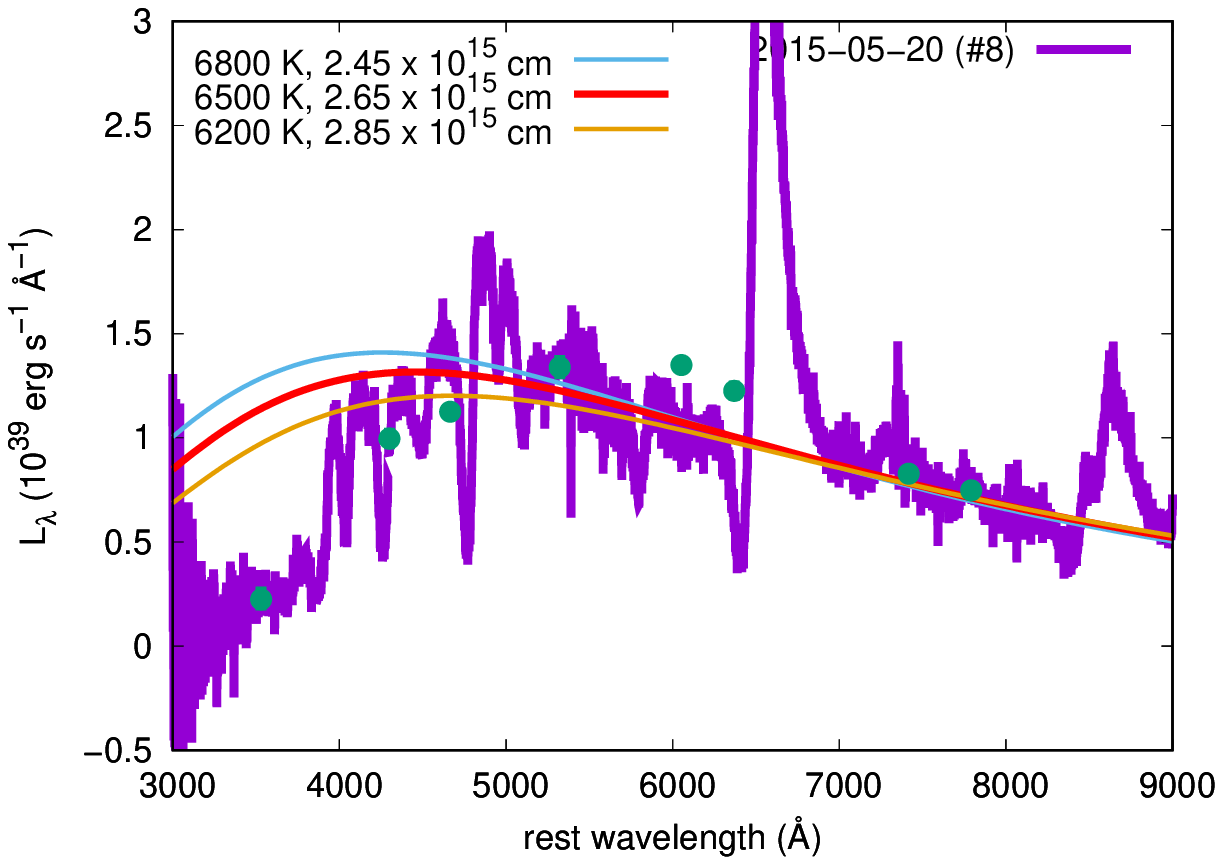} 
\includegraphics[width=0.66\columnwidth]{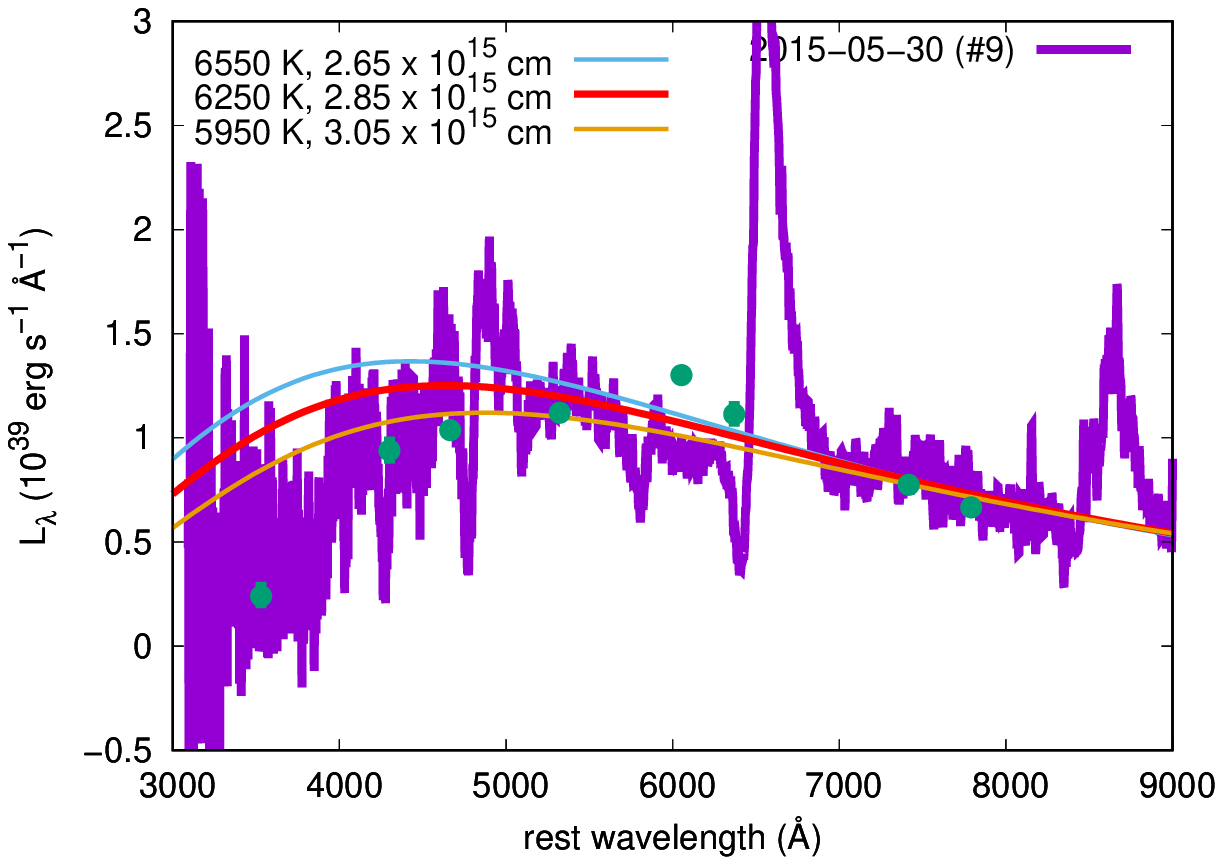} 
\includegraphics[width=0.66\columnwidth]{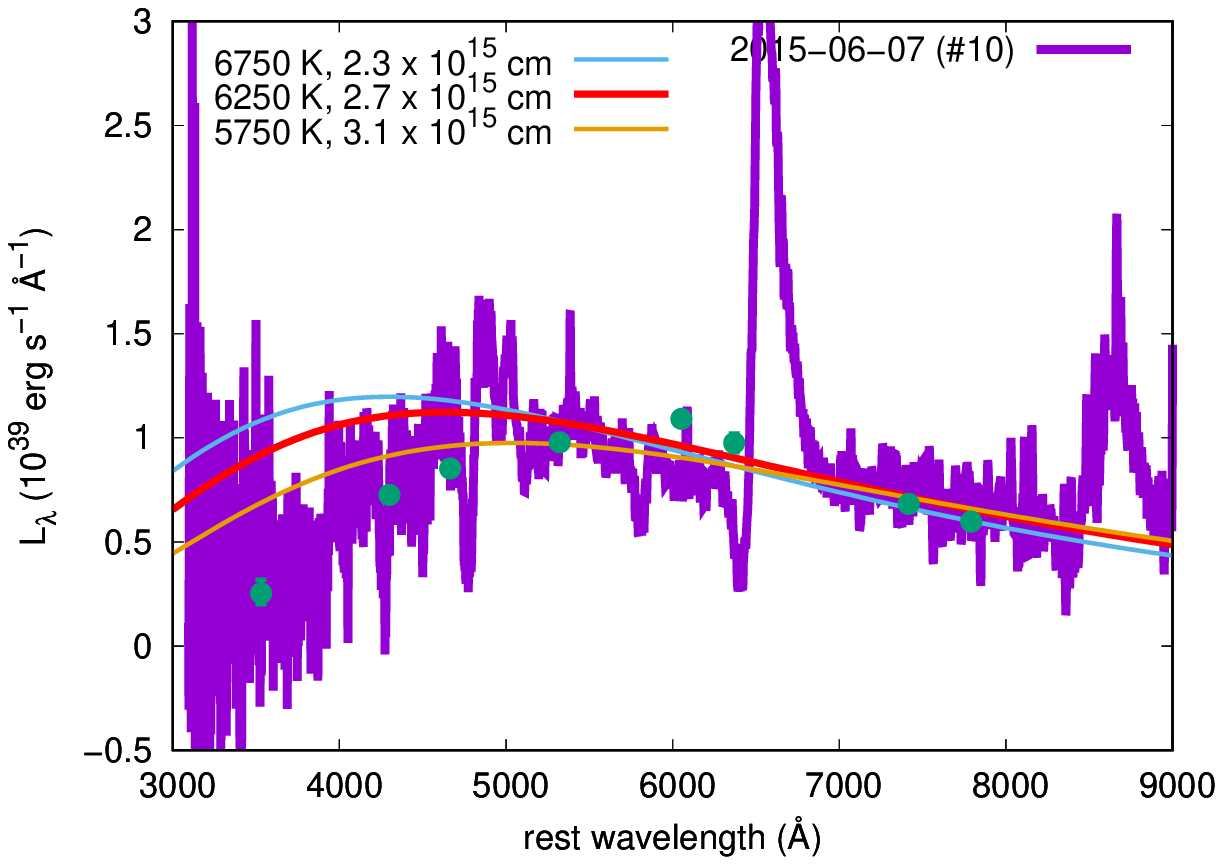} 
\includegraphics[width=0.66\columnwidth]{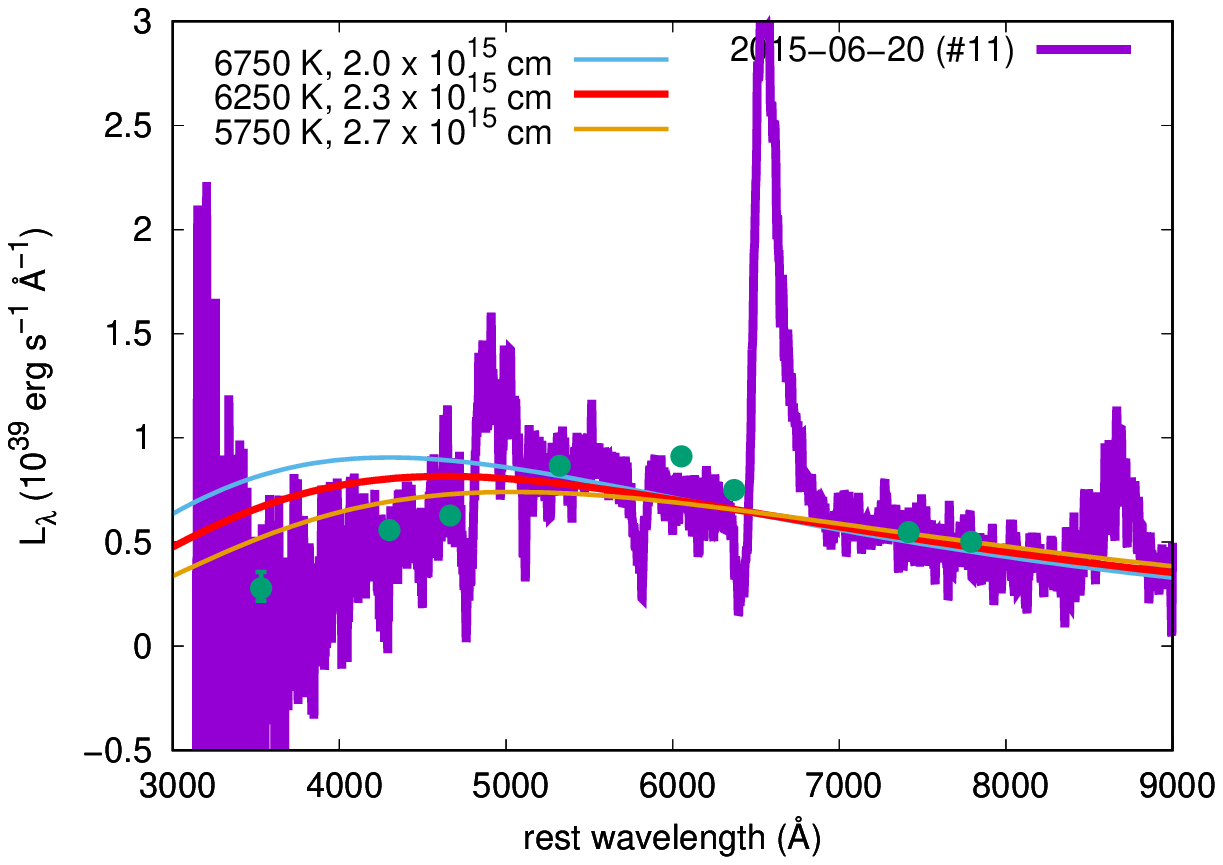} 
\includegraphics[width=0.66\columnwidth]{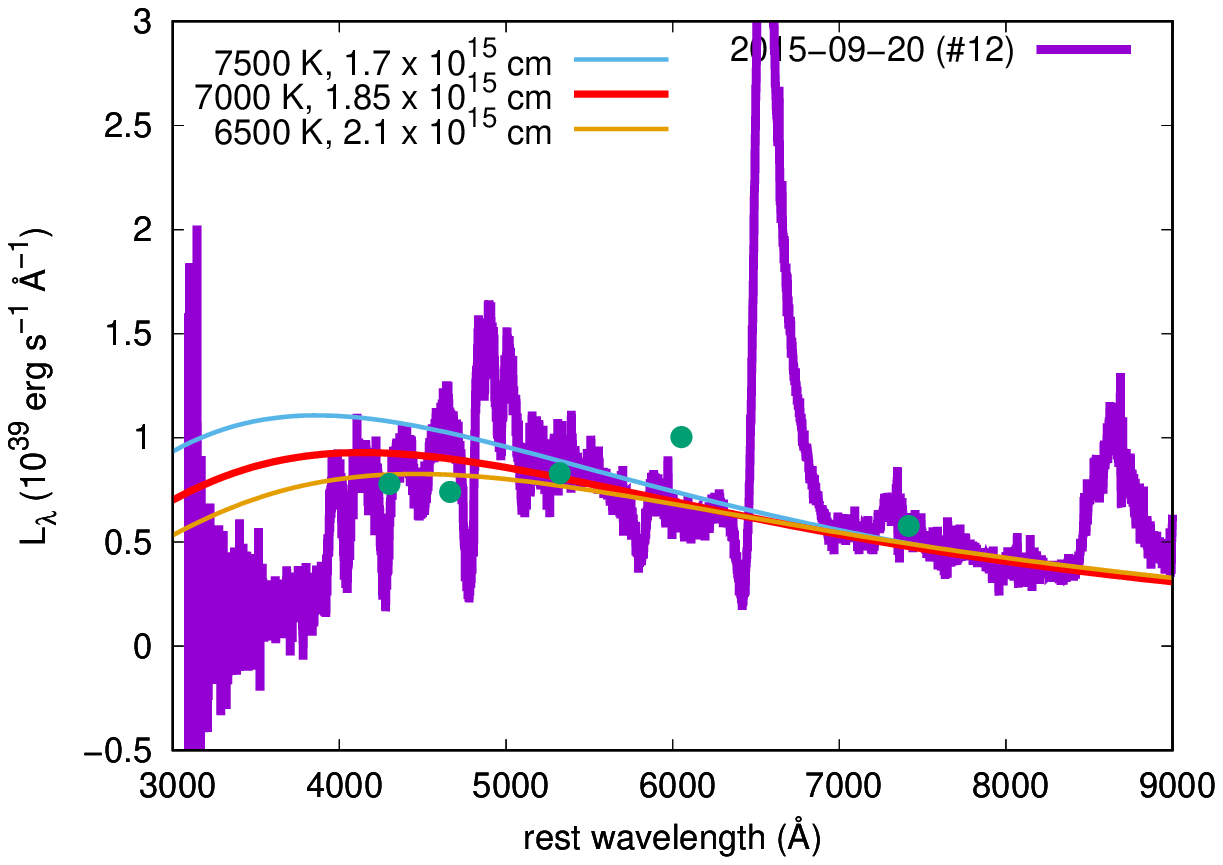} 
\includegraphics[width=0.66\columnwidth]{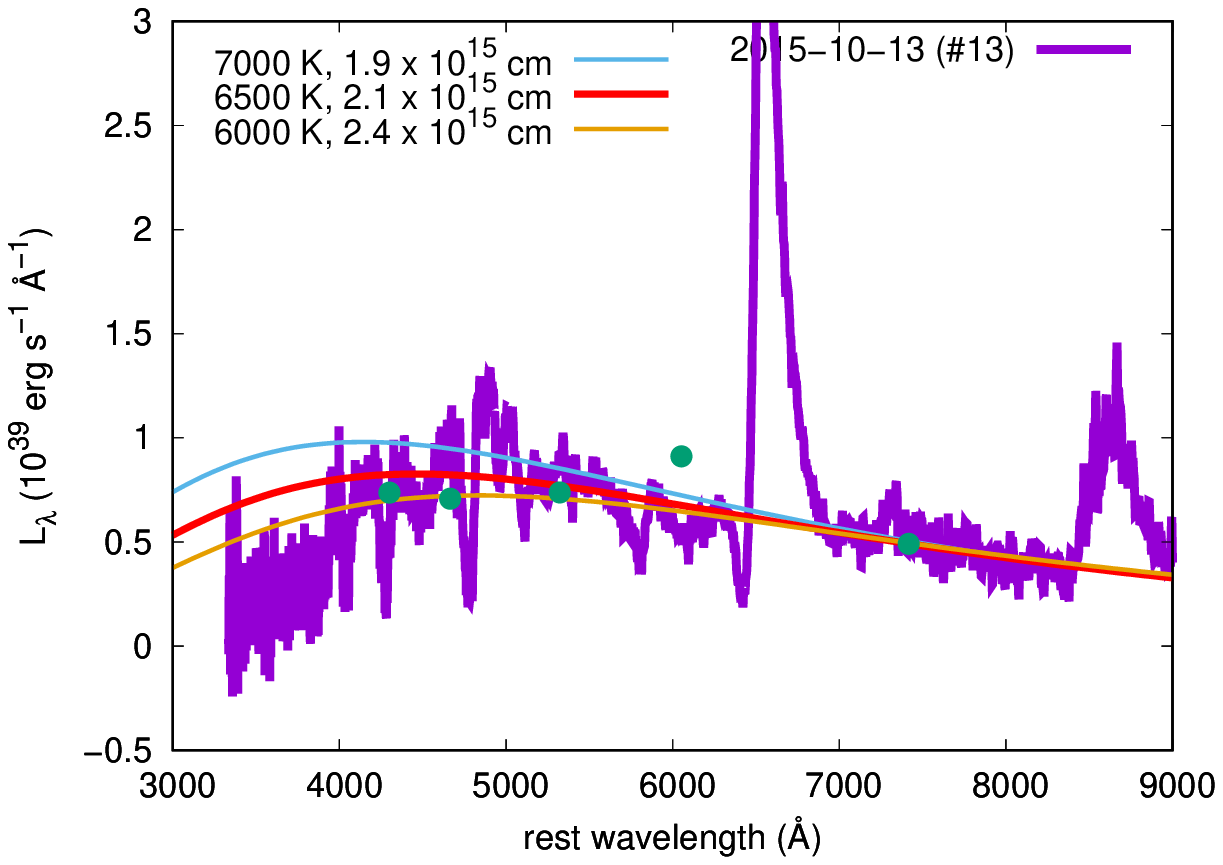} 
\includegraphics[width=0.66\columnwidth]{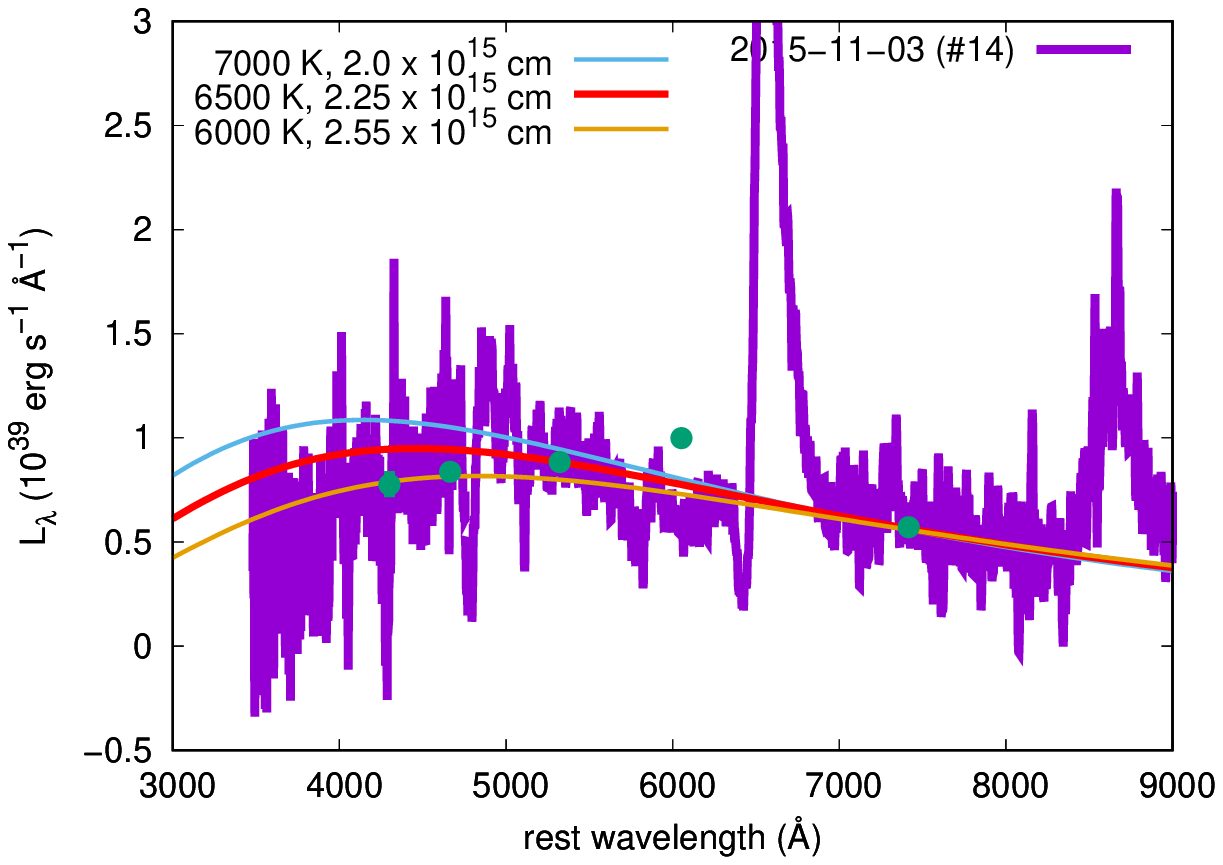} 
\includegraphics[width=0.66\columnwidth]{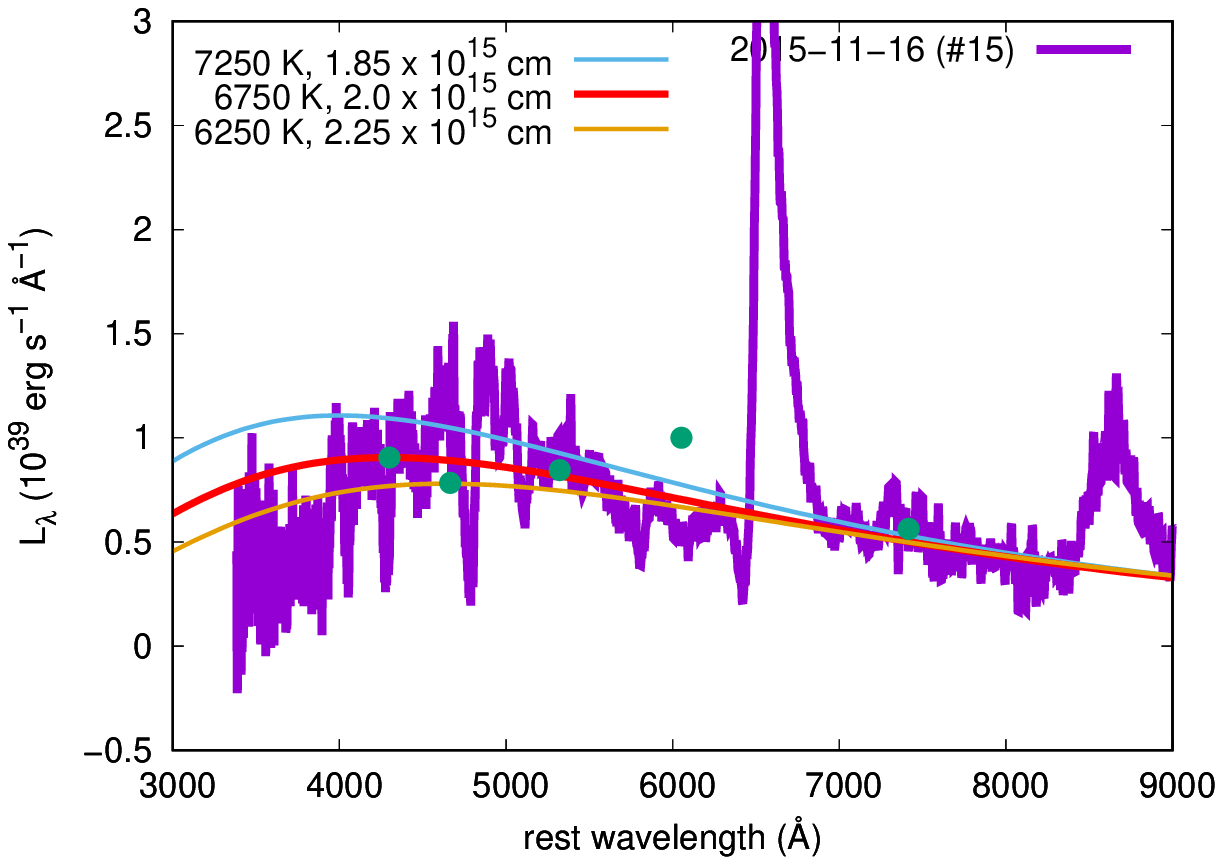} 
\includegraphics[width=0.66\columnwidth]{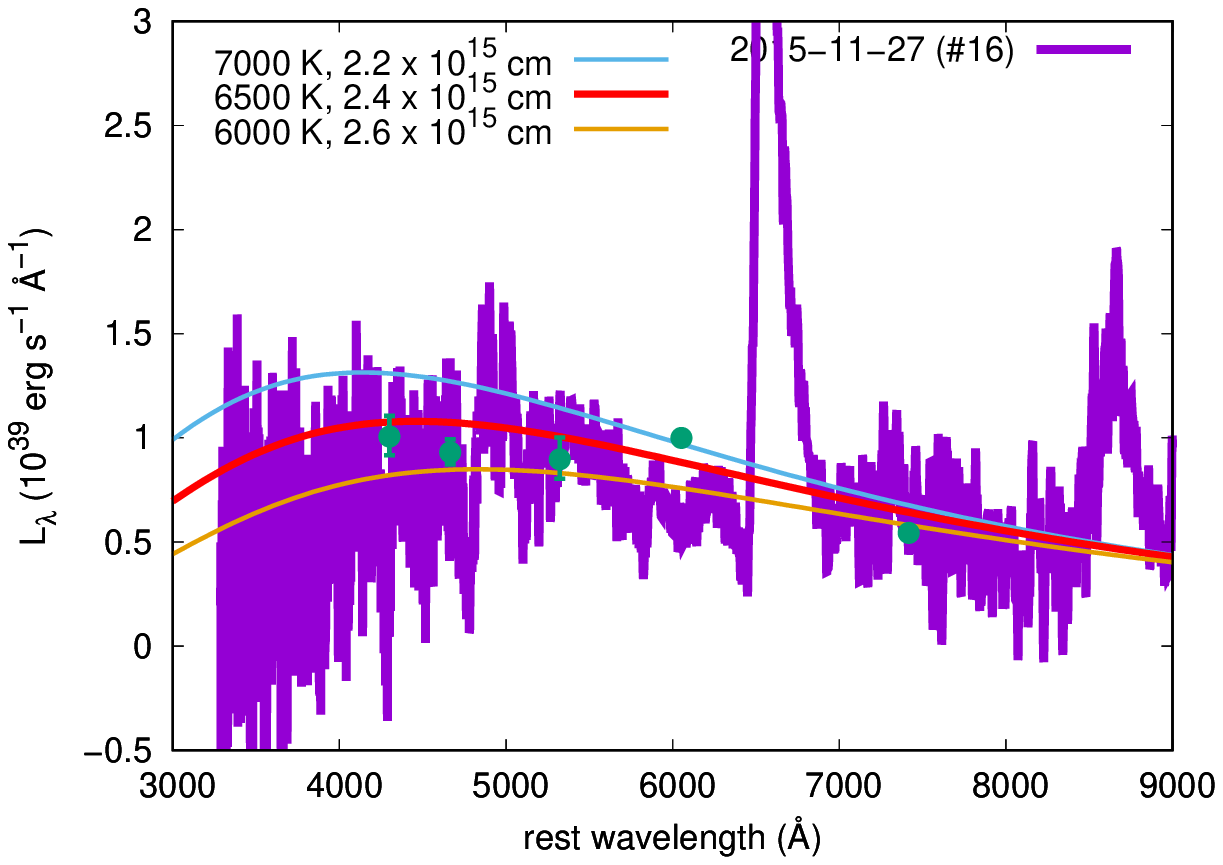} 
\includegraphics[width=0.66\columnwidth]{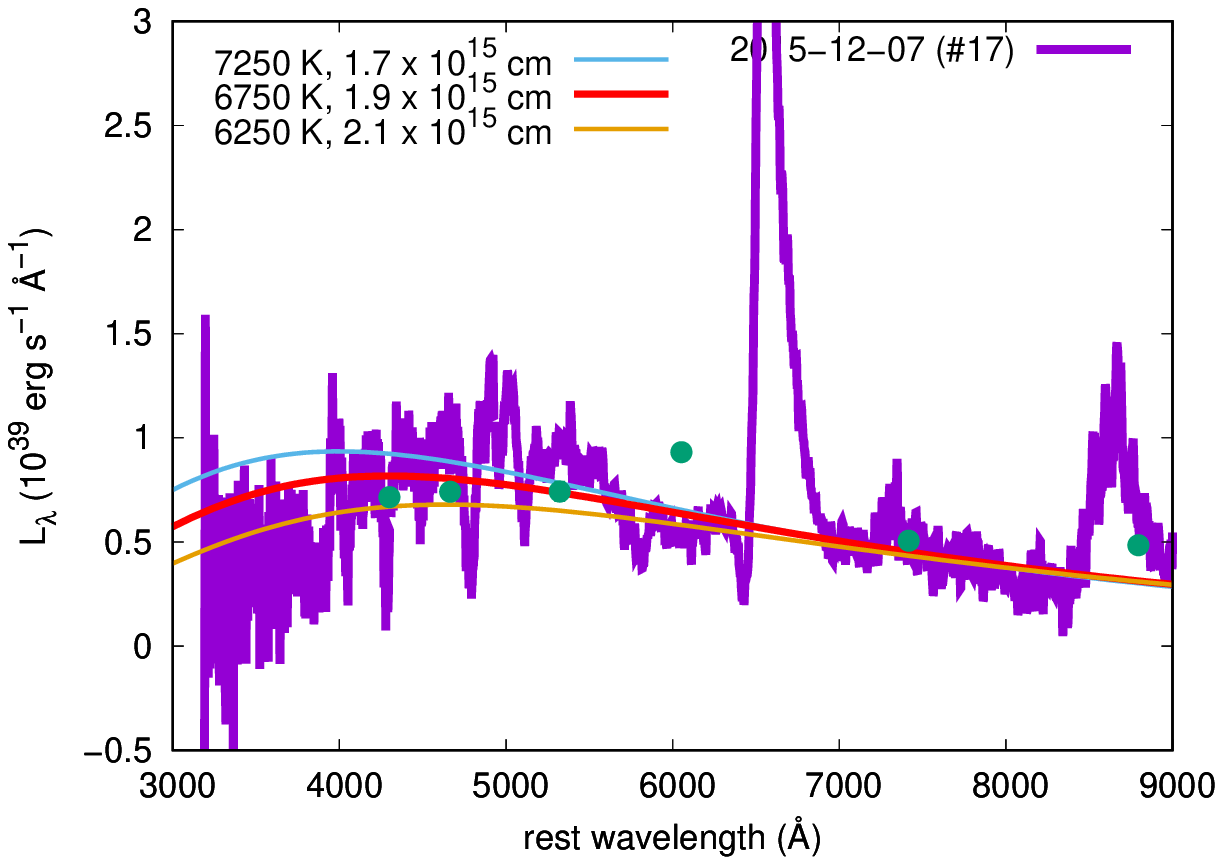} 
\includegraphics[width=0.66\columnwidth]{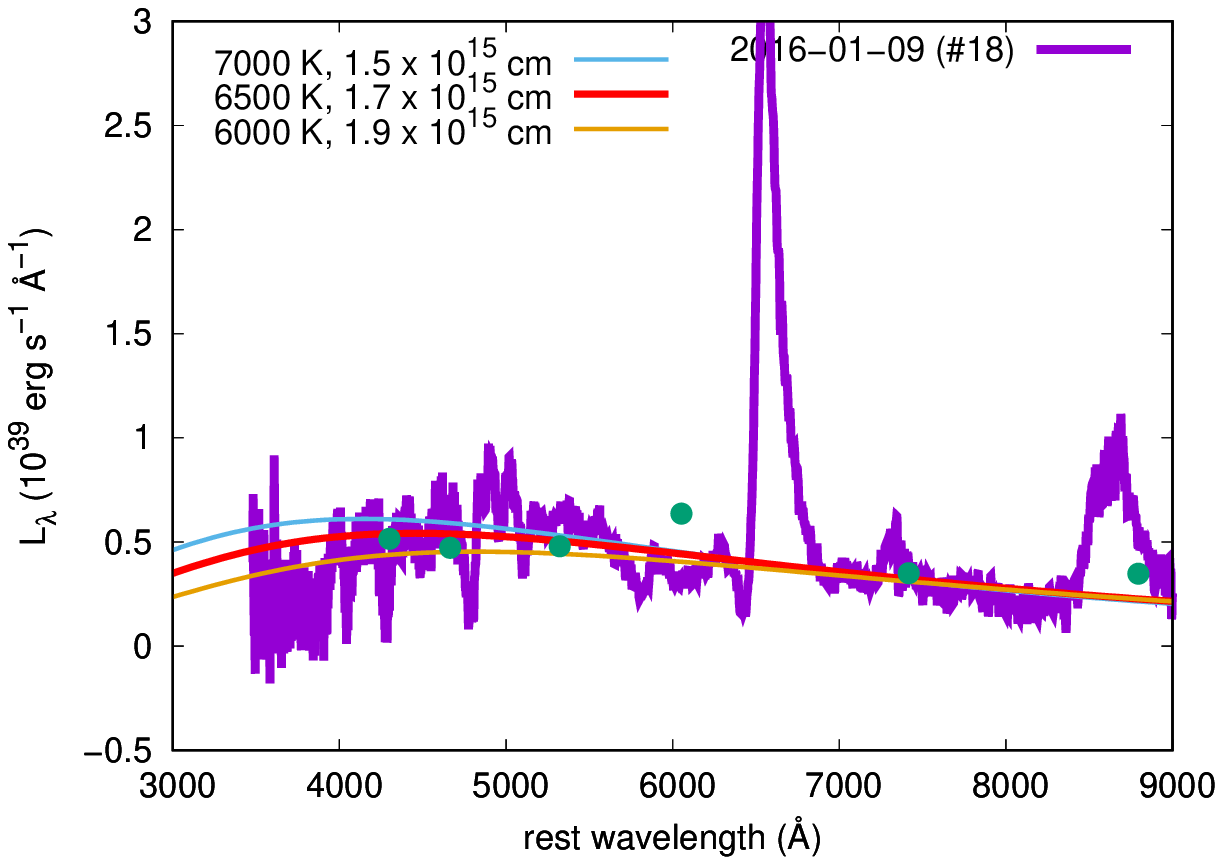} 
\caption{
Spectra and photometry (green circles) of iPTF14hls used for the blackbody fitting and their fitting results.
}
\label{fig:bbfits}
\end{figure*}

We present all the results of the blackbody fitting used in the paper in Fig.~\ref{fig:bbfits}.

\bsp	
\label{lastpage}
\end{document}